\documentclass[journal]{IEEEtran}
\usepackage{cite}
\usepackage{amssymb}
\usepackage{amsmath}
\usepackage{amssymb}
\usepackage{xspace}
\usepackage{graphicx}
\usepackage{subeqnarray}
\usepackage{cases}
\usepackage[ruled]{algorithm2e} 
\usepackage{stfloats}
\usepackage{subfigure}
\newtheorem{remark}{Remark}

\ifCLASSINFOpdf
\else
\fi

\begin{document}

\title{Generalized Minimum Error Entropy for Adaptive Filtering}

\author{Jiacheng He,
        Gang Wang,
        Bei Peng,
        ZhenYu Feng,
        and~Kun Zhang
\thanks{xxx was with the Department
of Electrical and Computer Engineering, Georgia Institute of Technology, Atlanta,
GA, 30332 USA e-mail: (see http://www.michaelshell.org/contact.html).}
\thanks{xxx and xxxx are with Anonymous University.}
\thanks{Manuscript received April 19, 2005; revised August 26, 2015.}}

\markboth{Journal of \LaTeX\ Class Files,~Vol.~14, No.~8, August~2015}%
{Shell \MakeLowercase{\textit{et al.}}: Bare Demo of IEEEtran.cls for IEEE Journals}

\maketitle

\begin{abstract}
Error entropy is a important nonlinear similarity measure, and it has received increasing attention in many practical applications. The default kernel function of error entropy criterion is Gaussian kernel function, however, which is not always the best choice. In our study, a novel concept, called generalized error entropy, utilizing the generalized Gaussian density (GGD) function as the kernel function is proposed. We further derivate the generalized minimum error entropy (GMEE) criterion, and a novel adaptive filtering called GMEE algorithm is derived by utilizing GMEE criterion.  The stability, steady-state performance, and computational complexity of the proposed algorithm are investigated. Some simulation indicate that the GMEE algorithm performs well in Gaussian, sub-Gaussian, and super-Gaussian noises environment, respectively. Finally, the GMEE algorithm is applied to acoustic echo cancelation and performs well.
\end{abstract}


\IEEEpeerreviewmaketitle

\section{Introduction} \label{Introduction}
\IEEEPARstart{T}{he} adaptive filtering algorithms have been extensively utilized in a variety of practical applications, such as active noise control (ANC) \cite{YIN2019138, SONG201969, TAN201529}, acoustic echo cancelation (AEC) \cite{pauline2020variable, WEN20191604}, and noise cancelation \cite{BAI2020430}. How to choose or construct an appropriate cost function is a key issue for adaptive filtering algorithm.

The distribution of the noise is an essential factor in determining the choice of the adaptive filtering cost function. Generally speaking, the common noise distributions are mainly divided into Gaussian, sub-Gaussian, and super-Gaussian distributions \cite{7426837, cichocki2002adaptive, WANG2018166}. Scholars, motivated by different noise distributions, proposed various optimization criteria (cost functions). The minimum mean square error (MMSE), as an important optimization criterion, plays a critical role when dealing with Gaussian noises. Some widely known algorithms based on MMSE criterion \cite{sayed2003fundamentals} are derived, such as least mean square (LMS) \cite{haykin2003least}, normalized LMS (NLMS) \cite{WANG201894}, variable step-size LMS (VSSLMS) \cite{4216658}, and fractional order modified least square (FOMLMS) \cite{CHENG201767}algorithms. Sub-Gaussian noises, such as uniform and binary noises, are also common noise distributions in the real environment. The least mean fourth (LMF) \cite{hubscher2003improved} and the least mean ${p}$-power \cite{pei1994least} algorithms outperform the LMS algorithm in sub-Gaussian noises environment. To address super-Gaussian noises (e.g., heavy-tailed impulse noises, Laplace, ${\alpha }$-stable, etc.), typical cost functions such as mixed-norm \cite{chambers1994least, chambers1997robust}, M-estimate cost \cite{zou2000least, chan2004recursive}, and correntropy \cite{ZHANG201612, CHEN2018318, WANG201988, PENG2017116} are utilized. Error entropy, as a widely known theory \cite{principe2010information}, takes higher order moments into account. Therefore, those algorithms founded on the minimum error entropy (MEE) criterion perform very well in impulsive (heavy-tailed) noises environment \cite{chen2018quantized}. Especially, the MEE criterion has already been successfully implemented in adaptive filtering \cite{chen2010mean, LI2020107534, wang2021adaptive, 8245817} and Kalman filter \cite{WANG2021107914}. 

The Gaussian kernel function is favored because of its smoothness and strict positive-definiteness, and it is always treated as the kernel function of error entropy. However, the default kernel function is not necessarily the best option \cite{7426837}. In our study, a new error entropy, called generalized error entropy, using the generalized Gaussian density (GGD) \cite{varanasi1989parametric} function as kernel function, is proposed. Moreover, we also propose a new learning criterion (or optimization criterion) called generalized minimum error entropy (GMEE) and a novel adaptive filtering based on GMEE criterion. Some important theoretical analysis of GMEE algorithm, such as stability, steady-state performance, and computational complexity, is investigated. Moreover, we compared the performance of the GMEE algorithm with some existing algorithms in respect of convergence speed, MSD, in Gaussian, sub-Gaussian and super-Gaussian noise environments, respectively. Finally, the new adaptive filtering is applied to AEC and performs well.

The organization of the rest of our work is presented below. The generalized error entropy is defined and described in Section \ref{section_definiton}. In Section \ref{Section_GMEE}, the proposed algorithm is derived based on the GMEE criterion. In Section \ref{Stalility_steady_state}, the stability, steady-state performance, and computational complexity of GMEE algorithm are investigated. In Section \ref{simulation}, some simulation examples and AEC experiment are presented to validate the theoretical results and the capabilities of the GMEE algorithm. The conclusion and acknowledgements are given in Section \ref{Conclusion} and \ref{Acknowledgements}, respectively.

\section{Definitions of Generalized Error Entropy and Quantized Generalized Error Entropy} \label{section_definiton}
\subsection{Generalized Error Entropy}
Renyi's seminal work on information theory is called Renyi’s ${\mu }$ entropy:
\begin{equation}\label{kekrvgolrhj1}
\begin{split}
{H_\mu }\left( e \right) = \frac{1}{{1 - \mu }}\log {V_\mu }\left( e \right).
\end{split}
\end{equation}
Here, ${\mu \left( {\mu  \ne 1,\mu  > 0} \right)}$ represent the order of Renyi’s entropy. The ${\mu }$ information potential (${\mu }$ IP) ${{V_\mu }\left( e \right)}$ of continuous variables is expressed as
\begin{equation}\label{zkhxkhe1jrfpED}
\begin{split}
{V_\mu }\left( e \right) = \int {{p^\mu }\left( e \right)} dx = {\text{E}}\left[ {{p^{\mu  - 1}}\left( e \right)} \right].
\end{split}
\end{equation}
Here ${p\left(  \cdot  \right)}$ denotes the probability density function (PDF) with respect to ${e}$, and ${{\text{E}}\left[  \cdot  \right]}$ stands for the expectation operator. In fact, PDF ${p\left( x \right)}$ is always estimated utilizing Parzen’s window strategy:
\begin{equation}\label{xkhiejxxksgmg}
\begin{split}
\hat p\left( x \right) = \frac{1}{L}\sum\limits_{i = 1}^L {{{\text{G}}_\sigma }} \left( {x - {e_i}} \right),
\end{split}
\end{equation}
where ${{{\text{G}}_\sigma }(x) = \left( {1/\sqrt {2\pi } \sigma } \right)\exp \left( { - {x^2}/2{\sigma ^2}} \right)}$ stands for the Gaussian kernel function, and ${\sigma }$ stands for kernel bandwidth, and ${\left\{ {{e_i}} \right\}_{i = 1}^L}$ are ${L}$ error samples. Combining (\ref{zkhxkhe1jrfpED}) and (\ref{xkhiejxxksgmg}), the estimation of the quadratic IP ${{V_2}\left( e \right)}$ can be obtained
\begin{equation}
\begin{split}
{\hat V_2}\left( e \right) = \frac{1}{L}\sum\limits_{i = 1}^L {\hat p} \left( e \right) = \frac{1}{{{L^2}}}\sum\limits_{i = 1}^L {\sum\limits_{j = 1}^L {{\operatorname{G} _\sigma }} } \left( {{e_i} - {e_j}} \right).
\end{split}
\end{equation}

The kernel function of conventional error entropy is always a Gaussian kernel function, however, it is not necessarily the best Option. The GGD function is a widely known extension of Gaussian density function \cite{varanasi1989parametric}:
\begin{equation}
\begin{split}
{\operatorname{G} _{\alpha ,\beta }}\left( e \right) = \frac{\alpha }{{2\beta \Gamma \left( {1/\alpha } \right)}}\exp \left( { - {{\left| {\frac{e}{\beta }} \right|}^\alpha }} \right).
\end{split}
\end{equation}

In our study, the GGD function is treated as the new kernel function of error entropy, and we define IP
\begin{equation}
\begin{split}\label{zhkuxikueajyibalz}
{V_{\alpha ,\beta }}\left( e \right) = \int {p_{\alpha ,\beta }^\alpha \left( e \right)de}  = \operatorname{E} \left[ {p_{\alpha ,\beta }^{\alpha  - 1}\left( e \right)} \right],
\end{split}
\end{equation}
where 
\begin{equation}
\begin{split}\label{eikjxkbetalphg}
{{\hat p}_{\alpha ,\beta }}\left( x \right) = \frac{1}{L}\sum\limits_{i = 1}^L {{\operatorname{G} _{\alpha ,\beta }}\left( {x - {e_i}} \right)} .
\end{split}
\end{equation}

In practical application, only a finite number of error set ${\left\{ {{e_i}} \right\}_1^L}$ can be obtained. Substituting (\ref{eikjxkbetalphg}) into (\ref{zhkuxikueajyibalz}) yields
\begin{equation}
\begin{split}\label{ValphabetamaoLFONE}
{\hat V_{\alpha ,\beta }}\left( e \right) = \frac{1}{L}\sum\limits_{i = 1}^L {{{\hat p}_{\alpha ,\beta }}} \left( {{e_i}} \right) = \frac{1}{{{L^2}}}\sum\limits_{i = 1}^L {\sum\limits_{j = 1}^L {{\operatorname{G} _{\alpha ,\beta }}} } \left( {{e_i} - {e_j}} \right).
\end{split}
\end{equation}
\subsection{Quantized Generalized Error Entropy}
According to (\ref{ValphabetamaoLFONE}), the IP can be calculated by using double summation method. This method causes a huge computational burden to get the IP, especially for large data sets. Therefore, on the basis of the generalized error entropy, we refer to previous studies \cite{8474935} to propose quantized generalized error entropy
\begin{equation}
\begin{split}\label{qgeecabcmCmM}
\begin{gathered}
  {{\hat V}_{\alpha ,\beta }}\left( e \right) = \frac{1}{L}\sum\limits_{i = 1}^L {{{\hat p}_{\alpha ,\beta }}} \left( {{e_i}} \right) \hfill \\
   \approx \hat V_{\alpha ,\beta }^Q\left( e \right) = \frac{1}{{{L^2}}}\sum\limits_{i = 1}^L {\sum\limits_{j = 1}^L {{\operatorname{G} _{\alpha ,\beta }}} } \left[ {{e_i} - {\text{Q}}\left[ {{e_j},\gamma } \right]} \right] \hfill \\
   = \frac{1}{{{L^2}}}\sum\limits_{i = 1}^L {\sum\limits_{h = 1}^H {{H_h}{\operatorname{G} _{\alpha ,\beta }}} } \left[ {{e_i} - {c_h}} \right] \hfill \\
   = \frac{1}{L}\hat p_{\alpha ,\beta }^Q\left( {{e_i}} \right). \hfill \\ 
\end{gathered} 
\end{split}
\end{equation}
Here, 
where quantization operator \cite{8474935} ${{\text{Q}}\left[ {{e_j},\gamma } \right] \in C}$ (${\gamma }$ is the quantization threshold) is used to obtain a codebook ${C = \left\{ {{c_1},{c_2}, \cdots {c_H} \in {\mathbb{R}^1}} \right\}}$ (in general ${H \leqslant L}$). ${{H_h}}$ is the number of error samples that are quantized to the code, ${\hat p_{\alpha ,\beta }^Q\left( e \right) = (1/L)\sum\nolimits_{h = 1}^H {{H_h}{{\text{G}}_{\alpha ,\beta }}(e - {c_h})} }$ is the PDF estimator based on the quantized error samples. Clearly, one has ${L = \sum\nolimits_{h = 1}^H {{H_h}} }$ and ${\int {\hat p_{\alpha ,\beta }^Q} \left( e \right)de = 1}$.

\begin{remark}
When ${H = L}$, one can obtain ${\hat p_{\alpha ,\beta }^Q\left( e \right) = {\hat p_{\alpha ,\beta }}\left( e \right)}$, which means that the quantized generalized error entropy reduces to the ordinary generalized error entropy.
\end{remark}

\section{Adaptive Filtering Base on GMEE} \label{Section_GMEE}
Consider a linear regression model:
\begin{equation}
\begin{split}
{d_n} = {\boldsymbol{w}}_s^{\text{T}}{{\boldsymbol{u}}_n} + {v_n},n = 1,2, \cdots ,N,
\end{split}
\end{equation}
where ${{{\boldsymbol{u}}_n} \in {\mathbb{R}^{M \times 1}}}$ denotes the input signal (a white sequence with zero mean and variance ${\sigma _u^2}$), ${n}$ is time point, ${{{\boldsymbol{w}}_s} \in {\mathbb{R}^{M \times 1}}}$ denotes the weight vector, and ${{v_n}}$ represents the additive noise. The error signal of the model is defined as
\begin{equation}
\begin{split}
{e_n} = {d_n} - {\boldsymbol{w}}_{n - 1}^{\text{T}}{{\boldsymbol{u}}_n}.
\end{split}
\end{equation}

\subsection{GMEE}
Generalized error entropy is similar to error entropy and can also serve as a criterion in estimation related problems. For linear adaptive filtering, under the generalized minimum error entropy (GMEE) criterion, the optimal weight vector of adaptive filtering can be calculated through minimizing the following function:
\begin{equation}
\begin{split}\label{jxjejiexbtedagssgml2}
{J_{{\text{GMEE }}}}\left( {{{\boldsymbol{w}}_n}} \right) = \mathop {\operatorname{argmin} }\limits_{{{\boldsymbol{w}}_n}} \frac{1}{{{L^2}}}\sum\limits_{i = n}^{n + L - 1} {\sum\limits_{j = n}^{n + L - 1} {{\operatorname{G} _{\alpha ,\beta }}} } \left( {{e_i} - {e_j}} \right).
\end{split}
\end{equation}
Under new optimization criterion (\ref{jxjejiexbtedagssgml2}), We derive a kind of random gradient-based novel adaptive filtering, and call it GMEE algorithm
\begin{equation}\label{12jdzuuijeejisinggau}
\begin{split}
\begin{gathered}
  {{\boldsymbol{w}}_{n + 1}} = {{\boldsymbol{w}}_n} + \eta \nabla {J_{{\text{GMEE }}}}\left( {{{\boldsymbol{w}}_n}} \right) \hfill \\
   = {{\boldsymbol{w}}_n} + \eta \frac{\alpha }{{{L^2}{\beta ^\alpha }}}\sum\limits_{i = n}^{n + L - 1} {\sum\limits_{j = n}^{n + L - 1} {\left[ \begin{gathered}
  {\operatorname{G} _{\alpha ,\beta }}\left( {{e_i} - {e_j}} \right){\left| {{e_i} - {e_j}} \right|^{\alpha  - 1}} \hfill \\
  \operatorname{sign} \left( {{e_i} - {e_j}} \right)\left( {{{\boldsymbol{u}}_i} - {{\boldsymbol{u}}_j}} \right) \hfill \\ 
\end{gathered}  \right]} }  \hfill \\
   = {{\boldsymbol{w}}_n} + \eta \frac{\alpha }{{{L^2}{\beta ^\alpha }}}{{\boldsymbol{U}}_n}\left( {{\boldsymbol{P}}_n^{\text{T}} - {\boldsymbol{Q}}_n^{\text{T}}} \right) \hfill \\ 
\end{gathered}   
\end{split}
\end{equation}
with 
\begin{subequations}
\begin{numcases}{}
{{\boldsymbol{P}}_n} = \left[ {\begin{array}{*{20}{c}}
  {{p_{n;n}}}&{{p_{n;n + 1}}}& \cdots &{{p_{n;n + L - 1}}} 
\end{array}} \right],\\ 
{p_{n;i}} = \sum\limits_{j = n}^{n + L - 1} {\left[ \begin{gathered}
  {\operatorname{G} _{\alpha ,\beta }}\left( {{e_i} - {e_j}} \right) \times  \hfill \\
  {\left| {{e_i} - {e_j}} \right|^{\alpha  - 1}}\operatorname{sign} \left( {{e_i} - {e_j}} \right) \hfill \\ 
\end{gathered}  \right],} \\
{q_{n;i}} = \sum\limits_{j = n}^{n + L - 1} {\left[ \begin{gathered}
  {\operatorname{G} _{\alpha ,\beta }}\left( {{e_j} - {e_i}} \right) \times  \hfill \\
  {\left| {{e_j} - {e_i}} \right|^{\alpha  - 1}}\operatorname{sign} \left( {{e_j} - {e_i}} \right) \hfill \\ 
\end{gathered}  \right]} \\
{{{\boldsymbol{U}}_n} = \left[ {\begin{array}{*{20}{c}}
  {{{\boldsymbol{u}}_n}}&{{{\boldsymbol{u}}_{n + 1}}}& \cdots &{{{\boldsymbol{u}}_{n + L - 1}}} 
\end{array}} \right].}
\end{numcases}
\end{subequations}
where ${\operatorname{sign} ( \cdot )}$ is SIGN function, ${\eta  > 0}$ is the step-size.
\begin{remark}
As a generalized minimum error entropy criterion, when ${\alpha  = 2}$, the GMEE algorithm reduces to
\begin{equation}
\begin{split}\label{MEEjkjiuujeebga}
{{\boldsymbol{w}}_{n + 1}} = {{\boldsymbol{w}}_n} + \eta \frac{2}{{{L^2}{\beta ^2}}}\sum\limits_{i = 1}^L {\sum\limits_{j = 1}^L {\left[ \begin{gathered}
  {\operatorname{G} _\beta }\left( {{e_i} - {e_j}} \right) \times  \hfill \\
  \left( {{e_i} - {e_j}} \right)\left( {{{\boldsymbol{u}}_i} - {{\boldsymbol{u}}_j}} \right) \hfill \\ 
\end{gathered}  \right]} } .
\end{split}
\end{equation}
It is obvious that (\ref{MEEjkjiuujeebga}) is the original MEE algorithm \cite{LI2020107534}.
\end{remark}

\begin{remark}
Various variants of the GMEE algorithm can be derived, such as novel recursive least squares (RLS) based on GMEE and variable kernel width GMEE algorithm.
\end{remark}

\subsection{QGMEE}
Similar to the derivation of (\ref{jxjejiexbtedagssgml2}) and (\ref{12jdzuuijeejisinggau}), we can obtain the cost function based on quantized generalized error entropy (\ref{qgeecabcmCmM})
\begin{equation}
\begin{split}
{J_{QGMEE{\text{ }}}}\left( {{{\boldsymbol{w}}_n}} \right) = \mathop {\operatorname{argmin} }\limits_{{{\boldsymbol{w}}_n}} \frac{1}{{{L^2}}}\sum\limits_{i = n}^{n + L - 1} {\sum\limits_{h = 1}^H {{H_h}{\operatorname{G} _{\alpha ,\beta }}\left[ {{e_i} - {c_h}} \right]} }  
\end{split}
\end{equation}
and an updated form of the weight vector
\begin{equation}
\begin{split}
\begin{gathered}
  {{\boldsymbol{w}}_{n + 1}} = {{\boldsymbol{w}}_n} + \eta \nabla {J_{QGMEE{\text{ }}}}\left( {{{\boldsymbol{w}}_n}} \right) \hfill \\
   = {{\boldsymbol{w}}_n} + \eta \frac{\alpha }{{{L^2}{\beta ^\alpha }}}\sum\limits_{i = n}^{n + L - 1} {\sum\limits_{h = 1}^H {\left[ \begin{gathered}
  {{\boldsymbol{u}}_i}{H_h}{\operatorname{G} _{\alpha ,\beta }}\left( {{e_i} - {c_h}} \right) \times  \hfill \\
  {\left| {{e_i} - {c_h}} \right|^{\alpha  - 1}}{\text{sign}}\left( {{e_i} - {c_h}} \right) \hfill \\ 
\end{gathered}  \right]} }  \hfill \\
   = {{\boldsymbol{w}}_n} + \eta \frac{\alpha }{{{L^2}{\beta ^\alpha }}}{{\boldsymbol{U}}_n}{\boldsymbol{\Lambda }}_n^{\text{T}}. \hfill \\ 
\end{gathered} 
\end{split}
\end{equation}
Here, ${{{\boldsymbol{\Lambda }}_n} = \left[ {\begin{array}{*{20}{c}}
  {{a_n}}&{{a_{n + 1}}}& \cdots &{{a_{n + L - 1}}} 
\end{array}} \right] \in {\mathbb{R}^{1 \times L}}}$ is a vector with 
\begin{equation}
\begin{split}
{a_i} = \sum\limits_{h = 1}^H {\left[ \begin{gathered}
  {H_h}{\operatorname{G} _{\alpha ,\beta }}\left( {{e_i} - {c_h}} \right) \times  \hfill \\
  {\left| {{e_i} - {c_h}} \right|^{\alpha  - 1}}{\text{sign}}\left( {{e_i} - {c_h}} \right) \hfill \\ 
\end{gathered}  \right]} .
\end{split}
\end{equation}
The computational burden of the GMEE algorithm can be effectively reduced by quantifying the set of errors ${\left\{ {{e_i}} \right\}_{i = 1}^L}$, and the computational complexity and performance of the proposed QGMEE are shown in Sections \ref{Stalility_steady_state} and \ref{simulation}, respectively.

\begin{remark}
Various variants of the GMEE algorithm can be derived, such as novel recursive least squares (RLS) based on GMEE and variable kernel width GMEE algorithm.
\end{remark}

\section{Stability and Steady-State Performance} \label{Stalility_steady_state}
In this part, some theoretical analysis of GMEE algorithm is implemented including stability, steady-state mean square behavior, and computational complexity. Before proceeding, two necessary assumptions are given as follows:

\emph{A1}: The element of priori errors ${{{\boldsymbol{\varepsilon }}_{a;n}}}$ and posteriori errors ${{{\boldsymbol{\varepsilon }}_{p;n}}}$ are independent of the noise.

\emph{A2}: The input signal and noise are uncorrelated at different instant.
\begin{equation}
\begin{split}
{\text{E}}\left[ {{{\boldsymbol{u}}_m}{\boldsymbol{u}}_n^{\text{T}}} \right] = \left\{ {\begin{array}{*{20}{l}}
  {\sigma _u^2{{\boldsymbol{I}}_M},m = n,} \\ 
  {0,m \ne n,} 
\end{array}} \right.
\end{split}
\end{equation}
\begin{equation}
\begin{split}
{\text{E}}\left[ {{\boldsymbol{u}}_m^{\text{T}}{{\boldsymbol{u}}_n}} \right] = \left\{ {\begin{array}{*{20}{l}}
  {M\sigma _u^2,m = n,} \\ 
  {0,m \ne n.} 
\end{array}} \right.
\end{split}
\end{equation}

\subsection{Stability Analysis}
The equation (\ref{12jdzuuijeejisinggau}) it is further formulated as below
\begin{equation}\label{15nTQjntpnUblea}
\begin{split}
{{{\boldsymbol{\tilde w}}}_{n + 1}} = {{{\boldsymbol{\tilde w}}}_n} - \eta \frac{\alpha }{{{L^2}{\beta ^\alpha }}}{{\boldsymbol{U}}_n}\left( {{\boldsymbol{P}}_n^{\text{T}} - {\boldsymbol{Q}}_n^{\text{T}}} \right),
\end{split}
\end{equation}
where ${{{\boldsymbol{\tilde w}}_n} = {{\boldsymbol{w}}_s} - {{\boldsymbol{w}}_n}}$, and priori and posteriori errors of the GMEE algorithm are defined by 
\begin{subequations}\label{defi16epsapns}
\begin{numcases}{}
{{\boldsymbol{\varepsilon }}_{a;n}} = {\boldsymbol{U}}_n^{\text{T}}{{{\boldsymbol{\tilde w}}}_n},\\ 
{{\boldsymbol{\varepsilon }}_{p;n}} = {\boldsymbol{U}}_n^{\text{T}}{{{\boldsymbol{\tilde w}}}_{n + 1}}.
\end{numcases}
\end{subequations}
and ${{{\boldsymbol{\varepsilon }}_{a;n}}{\text{ = }}\left[ {\begin{array}{*{20}{c}}
  {{{\boldsymbol{e}}_{a;n}}}&{{{\boldsymbol{e}}_{a;n + 1}}}& \cdots &{{{\boldsymbol{e}}_{a;n + L - 1}}} 
\end{array}} \right]}$.

Some expressions can obtain for simplicity
\begin{subequations}
\begin{numcases}{}
{{\boldsymbol{D}}_n} = {\left[ {\begin{array}{*{20}{c}}
  {{d_n}}&{{d_{n + 1}}}& \cdots &{{d_{n + L - 1}}} 
\end{array}} \right]^{\text{T}}},\\ 
{{\boldsymbol{V}}_n} = {\left[ {\begin{array}{*{20}{c}}
  {{v_n}}&{{v_{n + 1}}}& \cdots &{{v_{n + L - 1}}} 
\end{array}} \right]^{\text{T}}},\\
\begin{gathered}
  {{\boldsymbol{\varepsilon }}_n} = {\left[ {\begin{array}{*{20}{c}}
  {{e_n}}&{{e_{n + 1}}}& \cdots &{{e_{n + L - 1}}} 
\end{array}} \right]^{\text{T}}} \hfill \\
   = {{\boldsymbol{D}}_n} - {\boldsymbol{U}}_n^{\text{T}}{{\boldsymbol{w}}_n}. \hfill \\ 
\end{gathered} \label{dnjuvtwndyljyej17c} 
\end{numcases}
\end{subequations}
Moreover, by combining the defining (\ref{defi16epsapns}) and (\ref{dnjuvtwndyljyej17c}), one can obtain 
\begin{equation}
\begin{split}
{{\boldsymbol{\varepsilon }}_n} = {{\boldsymbol{\varepsilon }}_{a;n}} + {{\boldsymbol{V}}_n}.
\end{split}
\end{equation}
Left multiply both sides of (\ref{15nTQjntpnUblea}) by ${{\boldsymbol{U}}_n^{\text{T}}}$, and one can obtain 
\begin{equation}
\begin{split}
{{\boldsymbol{\varepsilon }}_{p;n}} = {{\boldsymbol{\varepsilon }}_{a;n}} - \eta \frac{\alpha }{{{L^2}{\beta ^\alpha }}}{\boldsymbol{U}}_n^{\text{T}}{{\boldsymbol{U}}_n}\left( {{\boldsymbol{P}}_n^{\text{T}} - {\boldsymbol{Q}}_n^{\text{T}}} \right).
\end{split}
\end{equation}
we can further get the following expression with assumption that matrix ${{{\boldsymbol{U}}_n^{\text{T}}{{\boldsymbol{U}}_n}}}$ is invertible
\begin{equation}\label{epsionapnj12UUT}
\begin{split}
\eta \frac{\alpha }{{{L^2}{\beta ^\alpha }}}\left( {{\boldsymbol{P}}_n^{\text{T}} - {\boldsymbol{Q}}_n^{\text{T}}} \right) = {\left( {{\boldsymbol{U}}_n^{\text{T}}{{\boldsymbol{U}}_n}} \right)^{ - 1}}\left( {{{\boldsymbol{\varepsilon }}_{a;n}} - {{\boldsymbol{\varepsilon }}_{p;n}}} \right).
\end{split}
\end{equation}
Substituting (\ref{epsionapnj12UUT}) into (\ref{15nTQjntpnUblea}), and one can obtain
\begin{equation}
\begin{split}
{{{\boldsymbol{\tilde w}}}_{n + 1}} = {{{\boldsymbol{\tilde w}}}_n} - {{\boldsymbol{U}}_n}{\left( {{\boldsymbol{U}}_n^{\text{T}}{{\boldsymbol{U}}_n}} \right)^{ - 1}}\left( {{{\boldsymbol{\varepsilon }}_{a;n}} - {{\boldsymbol{\varepsilon }}_{p;n}}} \right).
\end{split}
\end{equation}
We then square both sides to get
\begin{equation}
\begin{split}
\begin{gathered}
  {\left\| {{{{\boldsymbol{\tilde w}}}_{n + 1}}} \right\|^2} = {\left[ {{{{\boldsymbol{\tilde w}}}_n} - {{\boldsymbol{U}}_n}{{\left( {{\boldsymbol{U}}_n^{\text{T}}{{\boldsymbol{U}}_n}} \right)}^{ - 1}}\left( {{{\boldsymbol{\varepsilon }}_{a;n}} - {{\boldsymbol{\varepsilon }}_{p;n}}} \right)} \right]^{\text{T}}} \times  \hfill \\
  \left[ {{{{\boldsymbol{\tilde w}}}_n} - {{\boldsymbol{U}}_n}{{\left( {{\boldsymbol{U}}_n^{\text{T}}{{\boldsymbol{U}}_n}} \right)}^{ - 1}}\left( {{{\boldsymbol{\varepsilon }}_{a;n}} - {{\boldsymbol{\varepsilon }}_{p;n}}} \right)} \right]. \hfill \\ 
\end{gathered}   
\end{split}
\end{equation}
This yields, after some straightforward manipulations, the relation
\begin{equation}\label{23jnnttPQuunntEw}
\begin{split}
\begin{gathered}
  {\text{E}}\left[ {{{\left\| {{{{\boldsymbol{\tilde w}}}_{n + 1}}} \right\|}^2}} \right] = {\text{E}}\left[ {{{\left\| {{{{\boldsymbol{\tilde w}}}_n}} \right\|}^2}} \right] - 2\eta \frac{\alpha }{{{L^2}{\beta ^\alpha }}}{\text{E}}\left[ {{\boldsymbol{\varepsilon }}_{a;n}^{\text{T}}\left( {{\boldsymbol{P}}_n^{\text{T}} - {\boldsymbol{Q}}_n^{\text{T}}} \right)} \right] \hfill \\
   + {\left( {\eta \frac{\alpha }{{{L^2}{\beta ^\alpha }}}} \right)^2}{\text{E}}\left[ {\left( {{{\boldsymbol{P}}_n} - {{\boldsymbol{Q}}_n}} \right){\boldsymbol{U}}_n^{\text{T}}{{\boldsymbol{U}}_n}\left( {{\boldsymbol{P}}_n^{\text{T}} - {\boldsymbol{Q}}_n^{\text{T}}} \right)} \right]. \hfill \\ 
\end{gathered} 
\end{split}
\end{equation}
From (\ref{23jnnttPQuunntEw}), when it meets the conditions ${{\text{E}}\left[ {{{\left\| {{{{\boldsymbol{\tilde w}}}_{n + 1}}} \right\|}^2}} \right] = {\text{E}}\left[ {{{\left\| {{{{\boldsymbol{\tilde w}}}_n}} \right\|}^2}} \right]}$, one can obtain
\begin{equation}
\begin{split} \label{eq24dhzhQPttnnUUTEAT}
\eta  \leqslant \frac{{2{L^2}{\beta ^\alpha }{\text{E}}\left[ {{\boldsymbol{\varepsilon }}_{a;n}^{\text{T}}\left( {{\boldsymbol{P}}_n^{\text{T}} - {\boldsymbol{Q}}_n^{\text{T}}} \right)} \right]}}{{\alpha {\text{E}}\left[ {\left( {{{\boldsymbol{P}}_n} - {{\boldsymbol{Q}}_n}} \right){\boldsymbol{U}}_n^{\text{T}}{{\boldsymbol{U}}_n}\left( {{\boldsymbol{P}}_n^{\text{T}} - {\boldsymbol{Q}}_n^{\text{T}}} \right)} \right]}}
\end{split}
\end{equation}
According to assumption A2, the equation (\ref{eq24dhzhQPttnnUUTEAT}) can be rewritten as 
\begin{equation}
\begin{split}
\eta  \leqslant \frac{{2{L^2}{\beta ^\alpha }{\text{E}}\left[ {{\boldsymbol{\varepsilon }}_{a;n}^{\text{T}}\left( {{\boldsymbol{P}}_n^{\text{T}} - {\boldsymbol{Q}}_n^{\text{T}}} \right)} \right]}}{{\alpha {\text{E}}\left[ {\left( {{{\boldsymbol{P}}_n} - {{\boldsymbol{Q}}_n}} \right)} \right]{\text{E}}\left[ {{\boldsymbol{U}}_n^{\text{T}}{{\boldsymbol{U}}_n}} \right]{\text{E}}\left[ {\left( {{\boldsymbol{P}}_n^{\text{T}} - {\boldsymbol{Q}}_n^{\text{T}}} \right)} \right]}}.
\end{split}
\end{equation}
According to assumption A2, it is easy to obtain
\begin{equation}\label{LIUsigma2Md}
\begin{split}
{\text{E}}\left[ {{\boldsymbol{U}}_n^{\text{T}}{{\boldsymbol{U}}_n}} \right] = M\sigma _u^2{{\boldsymbol{I}}_L},
\end{split}
\end{equation}
 then one can obtain
\begin{equation}
\begin{split}\label{eq26v2nQTpjpnEQNPAmSIG}
\eta  \leqslant \frac{{2{L^2}{\beta ^\alpha }{\text{E}}\left[ {{\boldsymbol{\varepsilon }}_{a;n}^{\text{T}}} \right]{\text{E}}\left[ {\left( {{\boldsymbol{P}}_n^{\text{T}} - {\boldsymbol{Q}}_n^{\text{T}}} \right)} \right]}}{{\alpha M\sigma _u^2{\text{E}}\left[ {\left( {{{\boldsymbol{P}}_n} - {{\boldsymbol{Q}}_n}} \right)} \right]{\text{E}}\left[ {\left( {{\boldsymbol{P}}_n^{\text{T}} - {\boldsymbol{Q}}_n^{\text{T}}} \right)} \right]}},
\end{split}
\end{equation}

When time point ${n \to \infty }$, one can get ${{e_n} \to {v_n}}$, then matrices ${{{{\boldsymbol{P}}_n}}}$ and ${{{{\boldsymbol{Q}}_n}}}$ can be written as
\begin{equation}
\begin{split}
\left\{ \begin{gathered}
  {{{\boldsymbol{\tilde P}}}_n} = \left[ {\begin{array}{*{20}{c}}
  {{{\tilde p}_{n;n}}}&{{{\tilde p}_{n;n + 1}}}& \cdots &{{{\tilde p}_{n;n + L - 1}}} 
\end{array}} \right], \hfill \\
  {{{\boldsymbol{\tilde Q}}}_n} = \left[ {\begin{array}{*{20}{c}}
  {{{\tilde q}_{n;n}}}&{{{\tilde q}_{n;n + 1}}}& \cdots &{{{\tilde q}_{n;n + L - 1}}} 
\end{array}} \right], \hfill \\ 
\end{gathered}  \right.
\end{split}
\end{equation}
with 
\begin{equation}\label{31pqbbvvij}
\begin{split}
\left\{ \begin{gathered}
  {{\tilde p}_{n;i}} = {\text{E}}\left[ {\sum\limits_{j = n}^{n + L - 1} {\left[ \begin{gathered}
  {\operatorname{G} _{\alpha ,\beta }}\left( {{v_i} - {v_j}} \right) \times  \hfill \\
  {\left| {{v_i} - {v_j}} \right|^{\alpha  - 1}}\operatorname{sign} \left( {{v_i} - {v_j}} \right) \hfill \\ 
\end{gathered}  \right]} } \right], \hfill \\
  {{\tilde q}_{n;i}} = {\text{E}}\left[ {\sum\limits_{j = n}^{n + L - 1} {\left[ \begin{gathered}
  {\operatorname{G} _{\alpha ,\beta }}\left( {{v_j} - {v_i}} \right) \times  \hfill \\
  {\left| {{v_j} - {v_i}} \right|^{\alpha  - 1}}\operatorname{sign} \left( {{v_j} - {v_i}} \right) \hfill \\ 
\end{gathered}  \right]} } \right]. \hfill \\ 
\end{gathered}  \right.
\end{split}
\end{equation}
When the system is ergodic in a general sense, (\ref{31pqbbvvij}) can be further written as 
\begin{equation}
\begin{split}
\left\{ \begin{gathered}
  {{\tilde p}_{n;i}} \approx \frac{1}{i}\sum\limits_{t = 1}^i {\sum\limits_{j = n}^{n + L - 1} {\left[ \begin{gathered}
  {\operatorname{G} _{\alpha ,\beta }}\left( {{v_t} - {v_j}} \right) \times  \hfill \\
  {\left| {{v_t} - {v_j}} \right|^{\alpha  - 1}}\operatorname{sign} \left( {{v_t} - {v_j}} \right) \hfill \\ 
\end{gathered}  \right]} } , \hfill \\
  {{\tilde q}_{n;i}} \approx \frac{1}{i}\sum\limits_{t = 1}^i {\sum\limits_{j = n}^{n + L - 1} {\left[ \begin{gathered}
  {\operatorname{G} _{\alpha ,\beta }}\left( {{v_j} - {v_t}} \right) \times  \hfill \\
  {\left| {{v_j} - {v_t}} \right|^{\alpha  - 1}}\operatorname{sign} \left( {{v_j} - {v_t}} \right) \hfill \\ 
\end{gathered}  \right]} } . \hfill \\ 
\end{gathered}  \right.
\end{split}
\end{equation}
It is obvious that if step-size ${\eta }$ satisfies (\ref{eq26v2nQTpjpnEQNPAmSIG}),
then the sequence of ${{\text{E}}\left[ {{{\left\| {{{{\boldsymbol{\tilde w}}}_n}} \right\|}^2}} \right]}$ is convergent.

Similar to the derivation above (The stability analysis of the QGMEE algorithm is very similar to that of the GMEE algorithm, so we present the convergence conditions of the QGMEE algorithm directly.), one can obtain the convergence condition of the QGMEE algorithm as the following form
\begin{equation}
\begin{split}   \label{aanANTEKKMS}
\eta  \leqslant \frac{{2{L^2}{\beta ^\alpha }{\text{E}}\left[ {{\mathbf{\varepsilon }}_{a;n}^{\text{T}}} \right]{\text{E}}\left[ {{\mathbf{\Lambda }}_n^{\text{T}}} \right]}}{{\alpha M\sigma _u^2{\text{E}}\left[ {{{\mathbf{\Lambda }}_n}} \right]{\text{E}}\left[ {{\mathbf{\Lambda }}_n^{\text{T}}} \right]}}
\end{split}
\end{equation}
with 
\begin{equation}
\begin{split}  \label{38yudyuHhgalb}
{\text{E}}\left[ {{a_i}} \right] \approx \frac{1}{i}\sum\limits_{t = 1}^i {\sum\limits_{h = 1}^H {\left[ \begin{gathered}
  {H_h}{\operatorname{G} _{\alpha ,\beta }}\left( {{v_t} - {v_h}} \right) \times  \hfill \\
  {\left| {{v_t} - {v_h}} \right|^{\alpha  - 1}}{\text{sign}}\left( {{v_t} - {v_h}} \right) \hfill \\ 
\end{gathered}  \right]} } ,
\end{split}
\end{equation}
where ${{v_h} = Q\left[ {{e_j},\gamma } \right],(h = 1,2, \cdots ,H).}$
 One can obtain an approximation of the critical value of the step size of the QGMEE algorithm by calculating Eqs. (\ref{aanANTEKKMS}) and (\ref{38yudyuHhgalb}).

\subsection{Steady-State Mean Square Performance}
To investigated the steady-state behavior of GMEE algorithm, we start from the averaged relation (\ref{23jnnttPQuunntEw}). It is assumed that the GMEE algorithm is stable, and it will reach a steady state when it satisfies that 
\begin{equation}
\begin{split}
\mathop {\lim }\limits_{n \to \infty } {\text{E}}\left[ {{{\left\| {{{{\boldsymbol{\tilde w}}}_{n + 1}}} \right\|}^2}} \right] = \mathop {\lim }\limits_{n \to \infty } {\text{E}}\left[ {{{\left\| {{{{\boldsymbol{\tilde w}}}_n}} \right\|}^2}} \right].
\end{split}
\end{equation}
One can further obtain from (\ref{23jnnttPQuunntEw})
\begin{equation}\label{27nnpqttuunnETPQnn}
\begin{split}
\begin{gathered}
  {\text{E}}\left[ {{\boldsymbol{\varepsilon }}_{a;n}^{\text{T}}} \right]{\text{E}}\left[ {\left( {{\boldsymbol{P}}_n^{\text{T}} - {\boldsymbol{Q}}_n^{\text{T}}} \right)} \right] \hfill \\
   = \frac{\eta }{2}\left( {\frac{\alpha }{{{L^2}{\beta ^\alpha }}}} \right){\text{E}}\left[ {{{\left( {{\boldsymbol{P}}_n^{\text{T}} - {\boldsymbol{Q}}_n^{\text{T}}} \right)}^{\text{T}}}{\text{E}}\left[ {{\boldsymbol{U}}_n^{\text{T}}{{\boldsymbol{U}}_n}} \right]\left( {{\boldsymbol{P}}_n^{\text{T}} - {\boldsymbol{Q}}_n^{\text{T}}} \right)} \right]. \hfill \\ 
\end{gathered} 
\end{split}
\end{equation}
Left multiplying both sides of (\ref{27nnpqttuunnETPQnn}) by left inverse of ${{\text{E}}\left[ {\left( {{{\boldsymbol{P}}_n} - {{\boldsymbol{Q}}_n}} \right)} \right]}$
\begin{equation}
\begin{split}
{\text{E}}\left[ {{{\left[ {\left( {{\boldsymbol{P}}_n^{\text{T}} - {\boldsymbol{Q}}_n^{\text{T}}} \right)\left( {{{\boldsymbol{P}}_n} - {{\boldsymbol{Q}}_n}} \right)} \right]}^{ - 1}}\left( {{\boldsymbol{P}}_n^{\text{T}} - {\boldsymbol{Q}}_n^{\text{T}}} \right)} \right],
\end{split}
\end{equation}
and we obtain
\begin{equation}\label{31EEzhongxnnpttjnnusedjg}
\begin{split}
\begin{gathered}
  {\text{E}}\left[ {{{\boldsymbol{\varepsilon }}_{a;n}}} \right] = \frac{\eta }{2}\left( {\frac{\alpha }{{{L^2}{\beta ^\alpha }}}} \right){\text{E}}\left[ \begin{gathered}
  {\left[ {\left( {{\boldsymbol{P}}_n^{\text{T}} - {\boldsymbol{Q}}_n^{\text{T}}} \right)\left( {{{\boldsymbol{P}}_n} - {{\boldsymbol{Q}}_n}} \right)} \right]^{ - 1}} \hfill \\
   \times \left( {{\boldsymbol{P}}_n^{\text{T}} - {\boldsymbol{Q}}_n^{\text{T}}} \right) \hfill \\ 
\end{gathered}  \right] \hfill \\
   \times {\text{E}}\left[ {{{\left( {{\boldsymbol{P}}_n^{\text{T}} - {\boldsymbol{Q}}_n^{\text{T}}} \right)}^{\text{T}}}{\text{E}}\left[ {{\boldsymbol{U}}_n^{\text{T}}{{\boldsymbol{U}}_n}} \right]\left( {{\boldsymbol{P}}_n^{\text{T}} - {\boldsymbol{Q}}_n^{\text{T}}} \right)} \right]. \hfill \\ 
\end{gathered} 
\end{split}
\end{equation}
Substituting (\ref{LIUsigma2Md}) into (\ref{31EEzhongxnnpttjnnusedjg}), one can obtain
\begin{equation}
\begin{split}
{\text{E}}\left[ {{{\boldsymbol{\varepsilon }}_{a;n}}} \right] = \frac{\eta }{2}\left( {\frac{\alpha }{{{L^2}{\beta ^\alpha }}}} \right)M\sigma _u^2{\text{E}}\left[ {\left( {{\boldsymbol{P}}_n^{\text{T}} - {\boldsymbol{Q}}_n^{\text{T}}} \right)} \right].
\end{split}
\end{equation}
Assuming that the proposed adaptive filter is stable, we can obtain the steady-state excess mean square error (EMSE)
\begin{equation}
\begin{split} \label{eq34xzhxnqjnQnPEnttp}
\begin{gathered}
  \mathop {\lim }\limits_{n \to \infty } {\text{E}}\left[ {{\boldsymbol{e}}_{a;n}^2} \right] \approx \frac{1}{L}{\text{tr}}\left( {{\text{E}}\left[ {{{\boldsymbol{\varepsilon }}_{a;n}}{\boldsymbol{\varepsilon }}_{a;n}^{\text{T}}} \right]} \right) \hfill \\
   = \frac{{{\eta ^2}{\alpha ^2}{M^2}\sigma _u^4}}{{4{L^5}{\beta ^{2\alpha }}}}{\text{tr}}\left( {{\text{E}}\left[ {\left( {{\boldsymbol{P}}_n^{\text{T}} - {\boldsymbol{Q}}_n^{\text{T}}} \right)} \right]{\text{E}}\left[ {\left( {{{\boldsymbol{P}}_n} - {{\boldsymbol{Q}}_n}} \right)} \right]} \right). \hfill \\ 
\end{gathered} 
\end{split}
\end{equation}
When time point ${n \to \infty }$, one can get ${{e_n} \to {v_n}}$, then matrices ${{{{\boldsymbol{P}}_n}}}$ and ${{{{\boldsymbol{Q}}_n}}}$ can be computed. Knowing the noise distribution, the theoretical value of the EMSE can be calculated by (\ref{eq34xzhxnqjnQnPEnttp}).

As shown in table \ref{tab:compt_complex}, we compare the computational complexities of the GMEE algorithm with several adaptive algorithms including LMS, LMF, and GMCC algorithms. The GMEE algorithm has a slightly higher computational complexity compared with LMS, LMF, and GMCC algorithms. In addition, we also display the computational complexity of the QGMEE algorithm in Table \ref{tab:compt_complex}. 

In general, the number of real-valued code words ${H}$ is much less than the length of the sliding window ${L}$, as it shown in Table 2. We can infer that the adoption of the quantization mechanism can effectively reduce the computational complexity of the GMEE algorithm, and which has only a minimal negative impact on the performance of the GMEE algorithm (detailed simulations are shown in Section \ref{simulation}).

\begin{table*}
\centering
\caption{Computational complexities of different algorithms} \label{tab:compt_complex}
\begin{tabular}{llll}
\hline
Algorithms & ${ \times / \div }$   & ${ + / - }$   & Exponentiation          \\ 
\hline
LMS        & ${2M + 1}$             & ${2M}$             & ${0}$             \\
LMF        & ${2M + 1}$             & ${2M}$             & ${1}$             \\
GMCC       & ${2M + 4}$             & ${2M + 1}$         & ${3}$             \\
GMEE       & ${2M + ML + 6{L^2} + 3}$   & ${2M + ML + 8{L^2}}$     & ${6{L^2} + 2}$        \\
QGMEE       & ${M + ML + 4HL + 3}$      & ${M + ML + 4HL}$         & ${3HL + 2}$           \\
\hline
\end{tabular}
\end{table*}

\section{Simulations} \label{simulation}
We show some simulations to verify the theoretical results, in this part, and verify the outstanding performance of the GMEE algorithm. In addition, the GMEE algorithm is applied to the AEC experiment to verify the practicality. The mean-square deviation (MSD) was utilized to measure the performance of the adaptive filtering, and ${{\text{MSD}} = {\text{E[}}||{{\boldsymbol{w}}_s} - {{\boldsymbol{w}}_n}||_2^2]}$. The weight vector ${{{\boldsymbol{w}}_n}}$ is ${{\rm{10}} \times 1}$ vector. The signal ${{\boldsymbol{u}}}$ is white Gaussian random sequences with covariance ${{\text{E\{ }}{\boldsymbol{u}}{{\boldsymbol{u}}^{\text{T}}}{\text{\}  = }}{{\boldsymbol{I}}_{10}}}$ and ${{\text{E\{ }}{{\boldsymbol{u}}^{\text{T}}}{\boldsymbol{u}}{\text{\}  = 10}}}$
\subsection{Steady-State Performance}
\begin{figure}
\centerline{\includegraphics[width=\columnwidth]{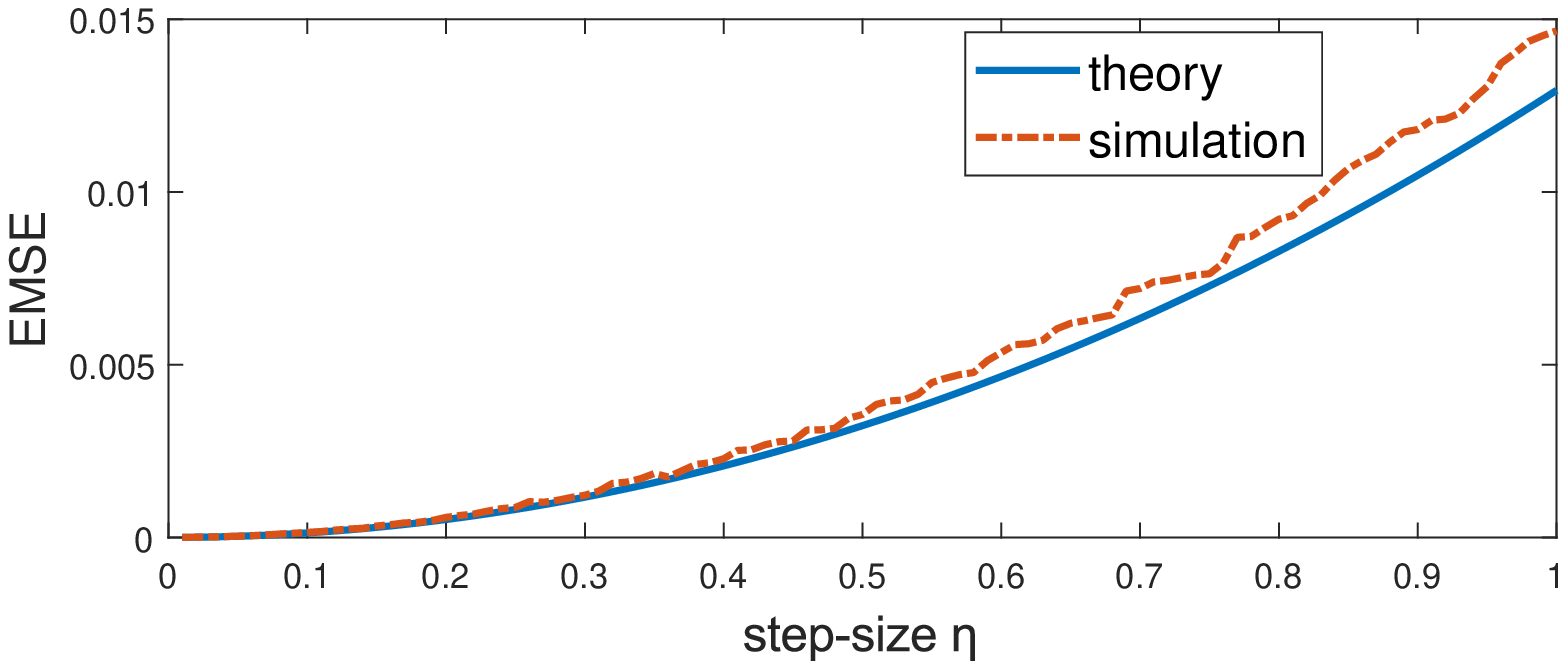}}
\caption{EMSE simulation theory}    \label{EMSE_simulation_theory}
\end{figure}

First, the theoretical and simulated steady-state value are investigated for the proposed GMEE. In this simulation, setting ${\eta  = 0.06}$, ${\alpha  = 2}$, ${\beta  = 1}$, and ${L = 10}$. The additive noise is mixed-Gaussian distribution with the form: ${{v_n} \sim 0.95N(0,0.01) + 0.05N(0,100)}$. To evaluate EMSEs, 100 independent simulations were performed, and 500000 iterations were performed to guarantee that the GMEE algorithm reaches a steady state in each simulation. Fig. \ref{EMSE_simulation_theory} presents the steady state of EMSEs at various step sizes, and we can obtain: 1) the steady state of EMSE increases with increasing step-size; 2) the steady state values of EMSEs calculated by simulation are very consistent with the theory values computed by (\ref{eq34xzhxnqjnQnPEnttp}), when the step-size is equal to a small value.
\subsection{Performance of GMEE}
Second, the performance of the GMEE algorithm is compared with several adaptive algorithms, such as the LMS, LMF, and GMCC \cite{7426837} algorithms. Four noise distributions are considered: a) Gaussian noise with zero-mean and unit variance; b) The normalized kurtosis of uniform sub-Gaussian noise is -1.25; c) Super-Gaussian noise is composed of a kind of mixed-Gaussian distribution with the form ${{v_n} \sim 0.95N(0,0.01) + 0.05N(0,100)}$; d) Super-Gaussian noise with zero mean Rayleigh distribution, which can be generated as ${{v_n} = {b_n}{R_n}}$ (${{R_n}}$ is zero mean Rayleigh noise), where ${{b_n}}$ is a Bernoulli process with ${{\rm{P}}\{ {b_i} = 1\}  = 0.3}$. In this subsection, the step-sizes that enables all of the algorithms to have the similar convergence speed are chosen. And 100 independent simulations were performed, 4000 iterations were run to evaluate MSD in every test. 

\begin{figure*}[htbp] 
\centering  
\subfigure[Gaussian noises.]{
\includegraphics[width=0.47\textwidth]{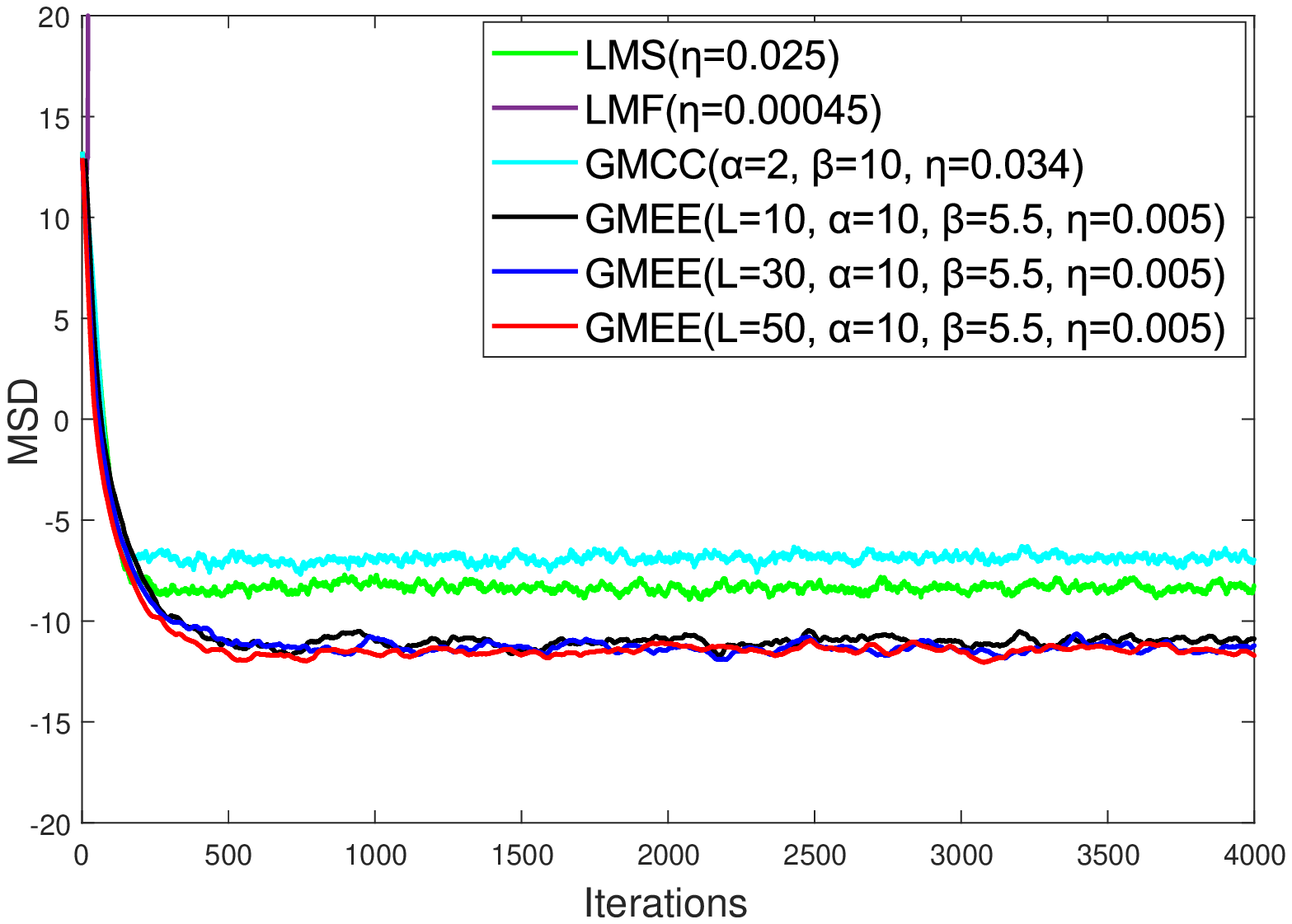}\label{fig_reslut_Gaussian}
}
\quad
\subfigure[Sub-Gaussian noises.]{
\includegraphics[width=0.47\textwidth]{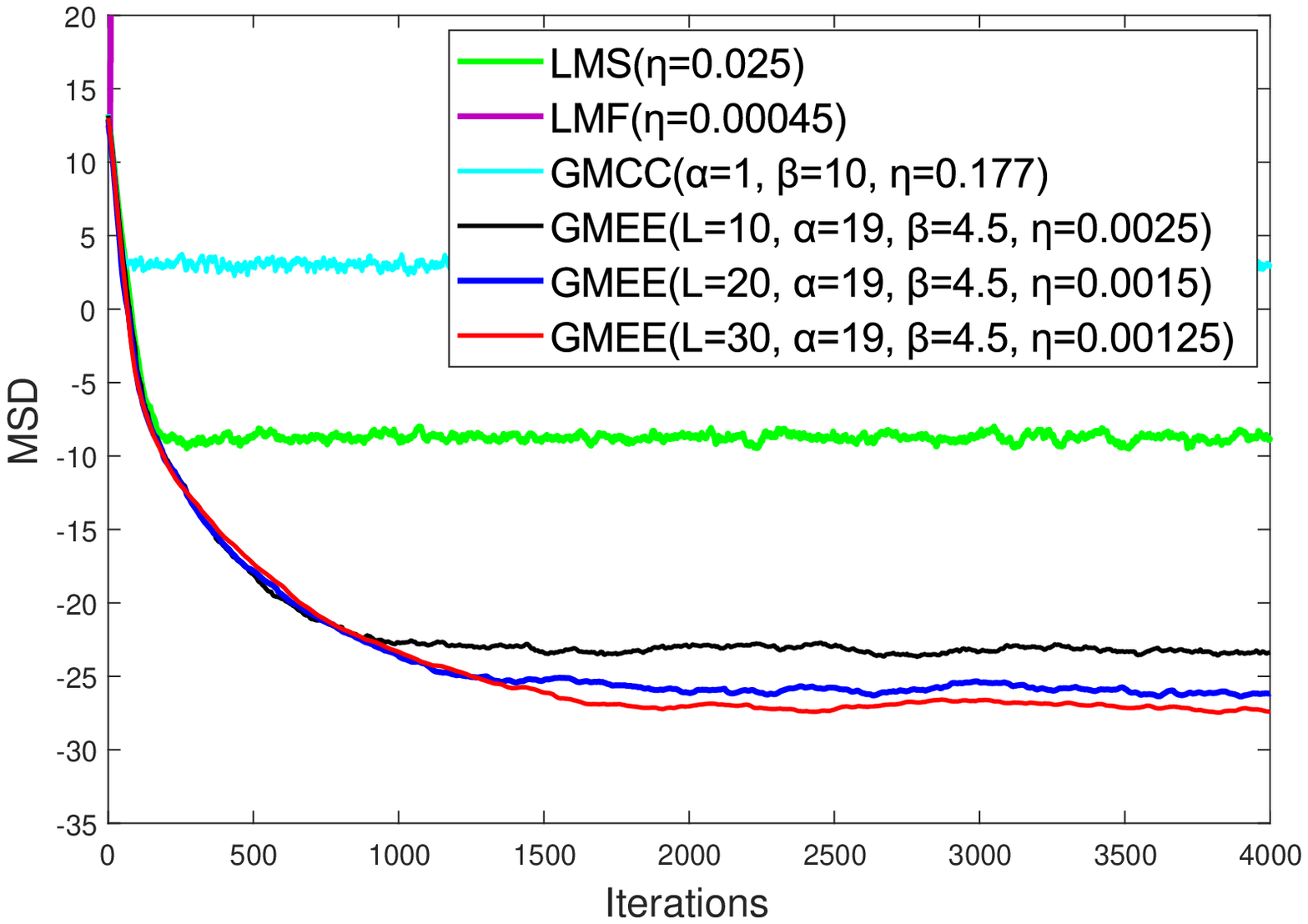}\label{fig_sub_Gaussian}
}
\quad
\subfigure[Mix-Gaussian noises.]{
\includegraphics[width=0.47\textwidth]{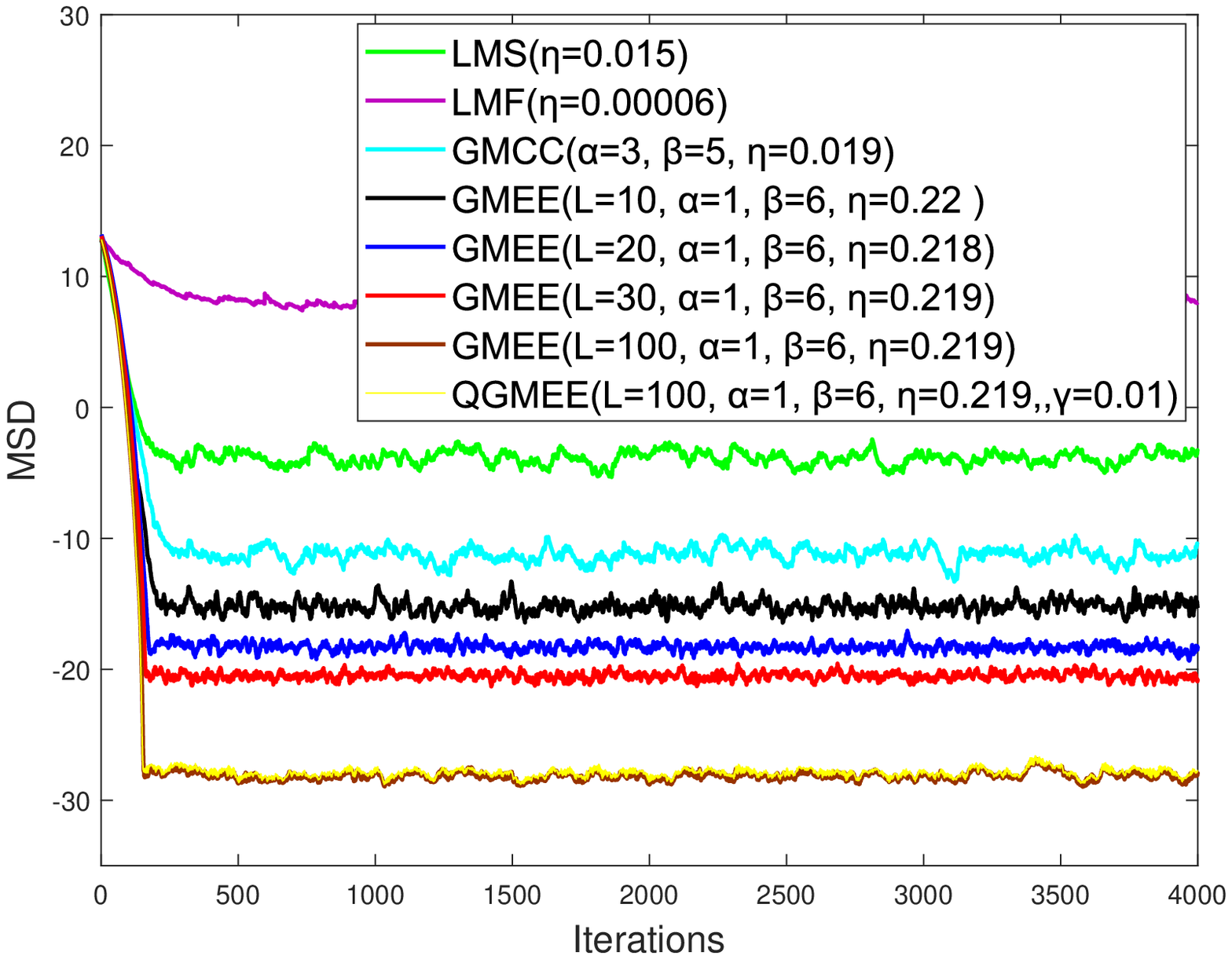}\label{fig_mix_Gaussian}
}
\quad
\subfigure[Rayleigh noise.]{
\includegraphics[width=0.47\textwidth]{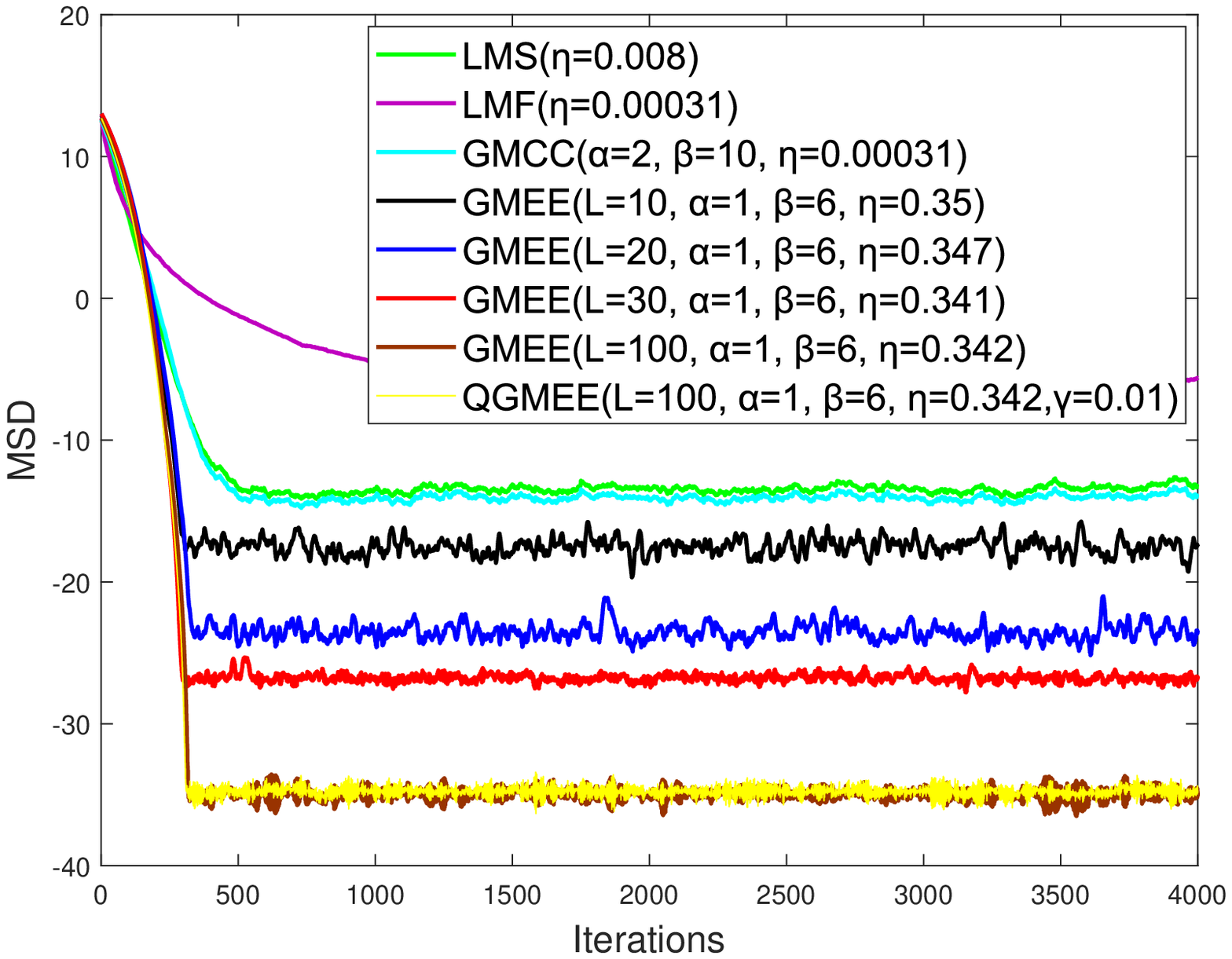}\label{fig_Rayleigh_noise}
}
\caption{Convergence curves under different noises}
\end{figure*}

The convergence curves of all algorithms in respect to MSD and final selected values of these parameters are displayed in fig. \ref{fig_reslut_Gaussian}, fig. \ref{fig_sub_Gaussian}, fig. \ref{fig_mix_Gaussian}, and fig. \ref{fig_Rayleigh_noise}. From simulation results we can observe: 1) Fig. \ref{fig_reslut_Gaussian} presents the steady error comparison under Gaussian noise, and one can get ${{\rm{GMEE < LMS < GMCC}}}$ (the LMF does not converge under this convergence speed); 2) The GMEE with proper parameters can outperform LMS, LMF, and GMCC algorithm under Gaussian, sub-Gaussian, and super-Gaussian noises. 3) The steady-state error of the GMEE algorithm decreases significantly as the number of error samples ${L}$ increases with the presence of sub-Gaussian noises and super-Gaussian noises, which also slightly increases the computational complexity of the GMEE. 

\begin{figure}
\centerline{\includegraphics[width=\columnwidth]{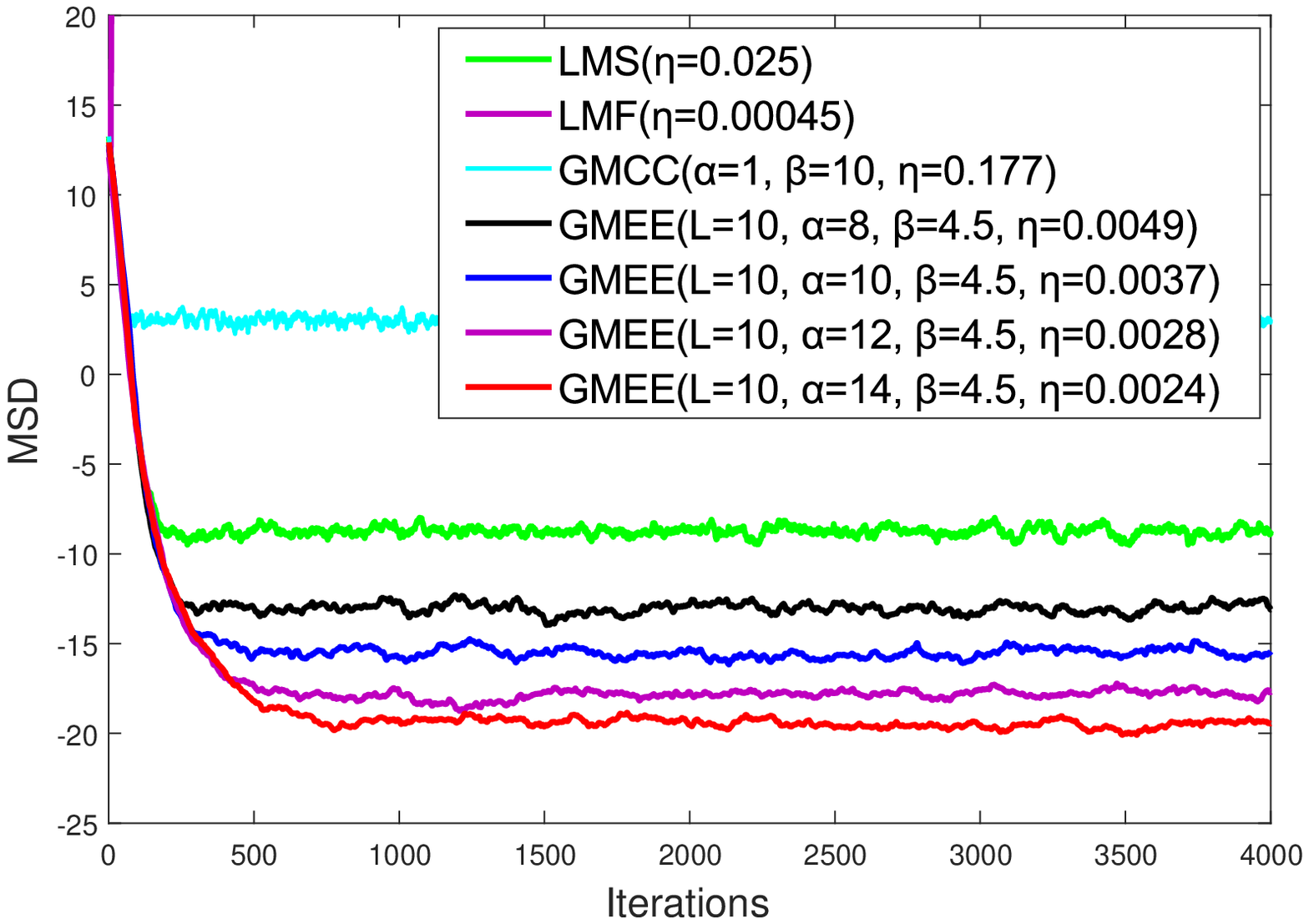}}
\caption{MSDs (dB) of the GMEE algorithm under different parameters (${\alpha }$) under sub-Gaussian noises.} \label{sub_Gaussian_alpha}
\end{figure}

To further investigate the effect of parameters ${\alpha }$ and ${\beta }$ with respect to the performance of the GMEE algorithm, we show the MSD (the initial convergence speed of GMEE with different ${\alpha }$ and ${\beta }$ is almost same) in Tables \ref{tab:performance alpha}, \ref{tab:performance beta} and fig. \ref{sub_Gaussian_alpha}. These four noise distributions are the same as the above noises and ${L = 10}$. It is easy to infer the influence of parameters ${\alpha }$ and ${\beta }$ with respect to the performance of the GMEE algorithm. There is no need for a detailed description of these results.

According to above simulations, one can infer that the GMEE algorithm perform well with Gaussian, super-Gaussian, and sub-Gaussian noise, espicilly in super- and sub- Gaussian noise. QGMEE algorithm have similar performance as GMEE with fewer calculations.

\begin{figure}
\centerline{\includegraphics[width=\columnwidth]{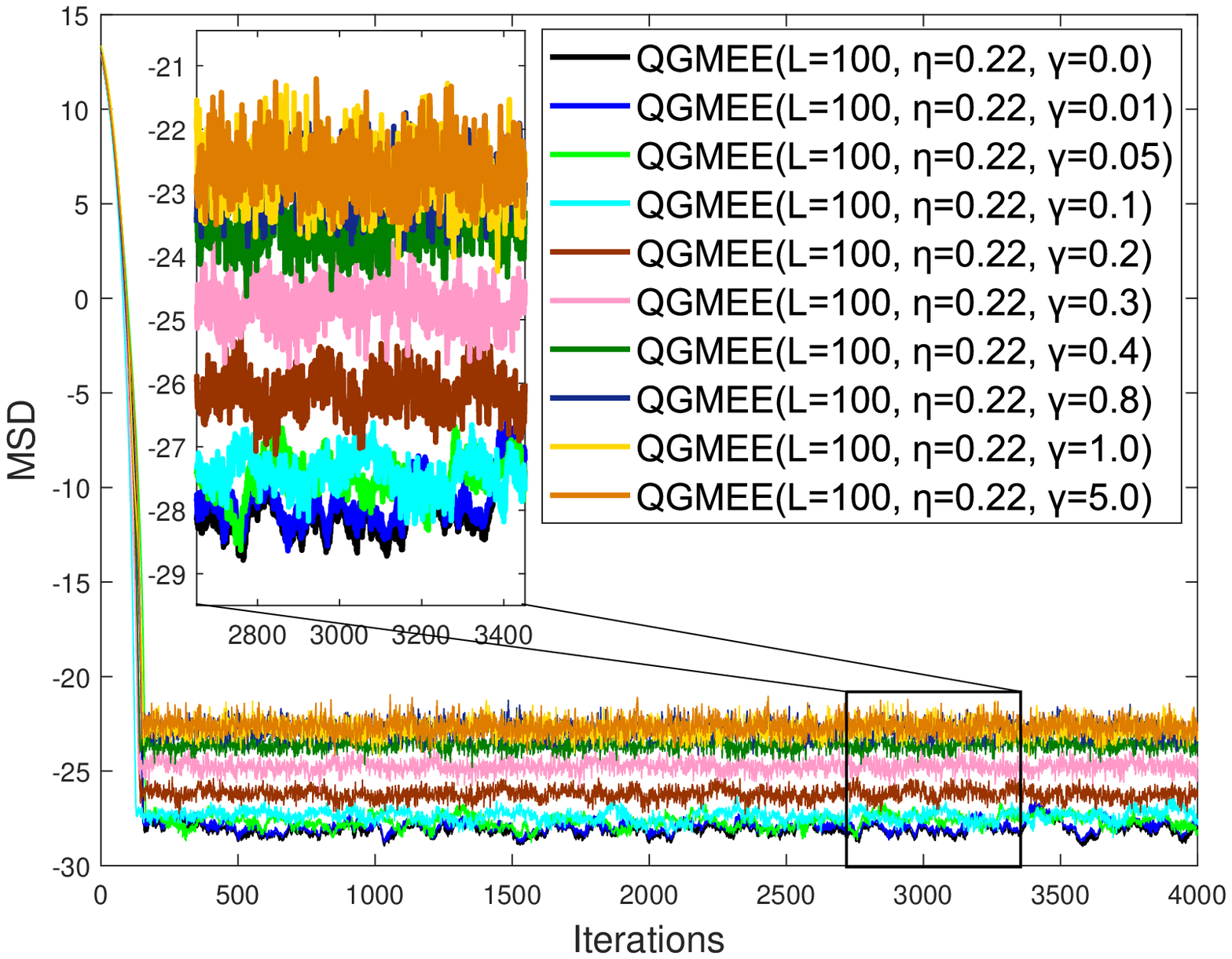}}
\caption{MSDs (dB) of the QGMEE algorithm under different  quantization  threshold (${\gamma }$) under mix-Gaussian noises.} \label{QGMEE_eta}
\end{figure}

\begin{table*}
  \centering
  \caption{Simulation steady-state error (dBs) of the GMEE with different ${\alpha}$.}
\label{tab:performance alpha}
\resizebox{\textwidth}{!}
{
\begin{tabular}{lllllllllll}
\hline
                    & ${\alpha  = 1}$      & ${\alpha  = 3}$     & ${\alpha  = 5}$      & ${\alpha  = 7}$      & ${\alpha  = 9}$      & ${\alpha  = 11}$     & ${\alpha  = 13}$      & ${\alpha  = 15}$     & ${\alpha  = 17}$     & ${\alpha  = 19}$     \\ \hline
Gaussian(${\beta  = 5.5}$)     & -1.82  & -6.87 & -9.55   & -9.95  & -10.11 & -11.21 & -10.73   & -10.20 & -9.64  & -8.11  \\
sub-Gaussian(${\beta  = 4.5}$) & -1.91   & -7.15 & -12.05 & -12.02 & -14.66 & -16.41 & -18.70   & -20.89 & -23.65 & -25.05 \\
mix-Gaussian(${\beta  = 6.0}$) & -15.95 & -7.96 & -3.19  & -1.53  & -0.94 & -0.29 & -0.030 & 0.44  & 0.66  & 0.88  \\
Rayleith(${\beta  = 6.0}$)     & -18.69 & -8.13 & -5.97   & -5.68  & -5.50  & -5.32  & -5.29   & -5.25   & -5.21  & -5.11   \\ \hline
\end{tabular}
}
\end{table*}

\begin{table*}
\caption{Simulation steady-state error (dBs) of the GMEE with different ${\beta}$.}
\label{tab:performance beta}
\resizebox{\textwidth}{!}
{
\begin{tabular}{lllllllllll}
\hline
                   & ${\beta  = 1}$   & ${\beta  = 3}$    & ${\beta  = 5}$    & ${\beta  = 7}$    & ${\beta  = 9}$    & ${\beta  = 11}$   & ${\beta  = 13}$   & ${\beta  = 15}$   & ${\beta  = 17}$   & ${\beta  = 19}$   \\ \hline
Gaussian(${\alpha  = 10}$ )     & 8.91  & -2.44  & -9.82  & -14.06 & -14.49 & -14.72 & -14.81 & -14.86 & -14.96 & -15.01 \\
sub-Gaussian(${\alpha  = 19}$ ) & 12.44 & 6.65   & -26.65 & -8.30  & -2.54  & 0.35   & 2.48   & 4.16   & 5.76   & 6.70   \\
mix-Gaussian(${\alpha  = 1}$ ) & -0.68 & -12.20 & -15.86 & -16.12 & -16.78 & -16.48 & -16.95 & -16.82 & -16.92 & -17.25 \\
Rayleith(${\alpha  = 1}$ )     & -2.17 & -15.90 & -17.39 & -18.62 & -19.35 & -19.93 & -20.30 & -20.41 & -20.53 & -20.55 \\ \hline
\end{tabular}
}
\end{table*}

\subsection{Application to AEC}
\begin{figure}
\centerline{\includegraphics[width=\columnwidth]{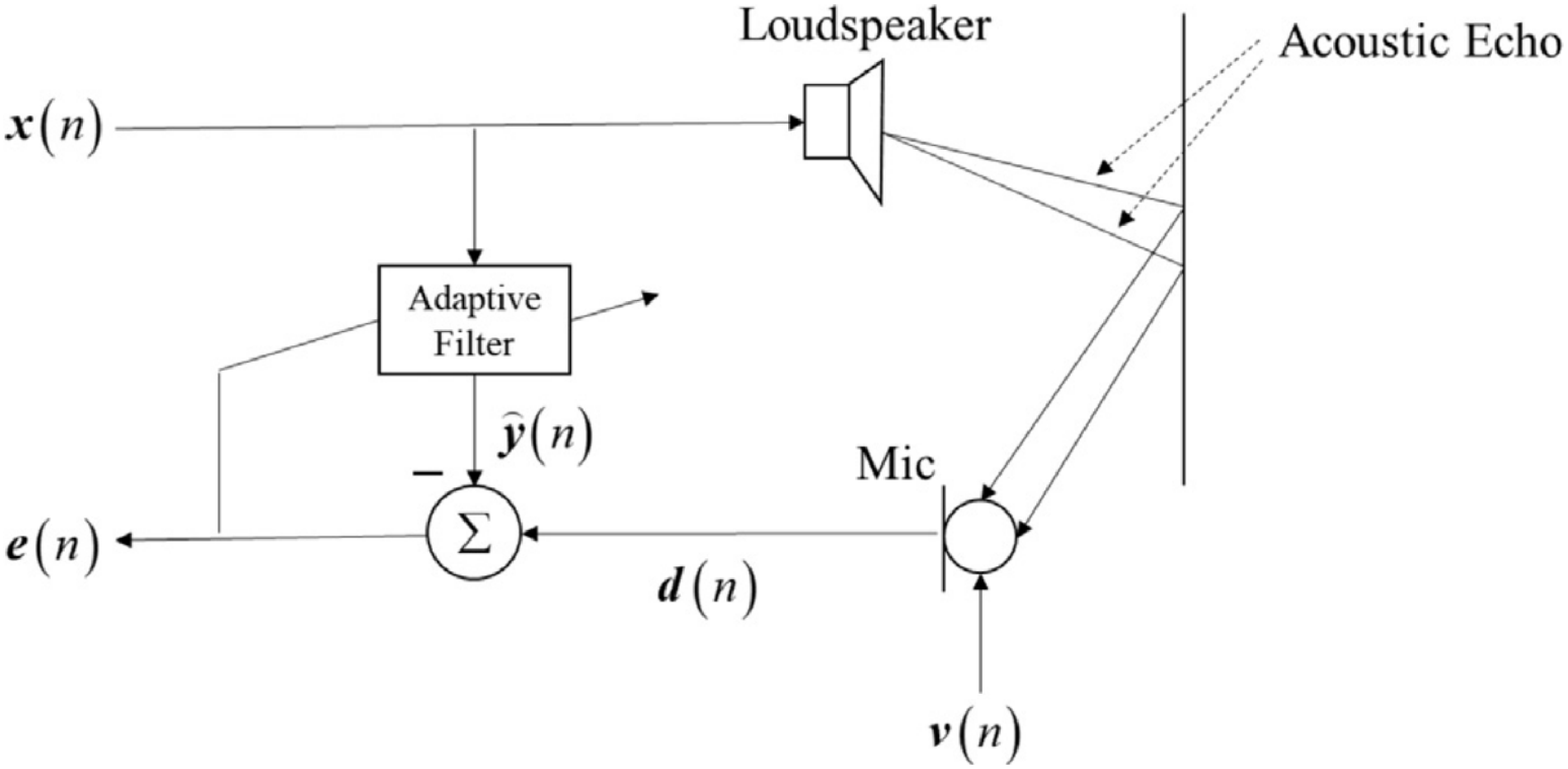}}
\caption{The schematic diagram of the AEC.} \label{sounds}
\end{figure}

In this part, the proposed GMEE algorithm is applied to AEC \cite{pauline2020variable} to validate the practical performance of the GMEE algorithm. As it illustrated in fig. \ref{sounds}, the microphone signal ${{\boldsymbol{d}}(n)}$ is the summation of desired signal ${{\boldsymbol{p}}(n)}$, echo signal ${{\boldsymbol{y}}(n)}$ generated by sound ${{\boldsymbol{x}}(n)}$, and noise signal ${{\boldsymbol{v}}(n)}$ with the form of ${{\boldsymbol{v}}(n) \sim 0.95N(0,0.001) + 0.05N(0,0.01)}$. The core of echo cancellation is to estimate the unknown function ${{\boldsymbol{\overset{\lower0.5em\hbox{$\smash{\scriptscriptstyle\frown}$}}{y} }}(n) = \Gamma \left[ {{\boldsymbol{x}}(n)} \right]}$ to eliminate echoes. However, due to the uncertainty and complexity of the reflection path, we simplify it to the linear form ${{\boldsymbol{\overset{\lower0.5em\hbox{$\smash{\scriptscriptstyle\frown}$}}{y} }}(n) = {{\boldsymbol{w}}^{\text{T}}}{\boldsymbol{x}}(n)}$, and the weight parameters can be estimated by adaptive filtering algorithms such as GMEE, LMS, and so on. This processed sound signal is ${{\boldsymbol{e}}(n) = {\boldsymbol{d}}(n) - {\boldsymbol{\overset{\lower0.5em\hbox{$\smash{\scriptscriptstyle\frown}$}}{y} }}(n)}$.

The performance of GMEE is compared with LMS and RLS algorithms in respect to echo return loss enhancement (ERLE), MSD, and convergence speed. We obtained these results through 500 Monte-Carlo experiments. The ERLE \cite{pauline2020variable} is defined as 
\begin{equation}
\begin{split}
{\text{ERLE}} = 10{\log _{10}}\frac{{{\text{power of far - end signal}}}}{{{\text{power of residual echo signal}}}}.
\end{split}
\end{equation}

\begin{figure}
\centerline{\includegraphics[width=\columnwidth]{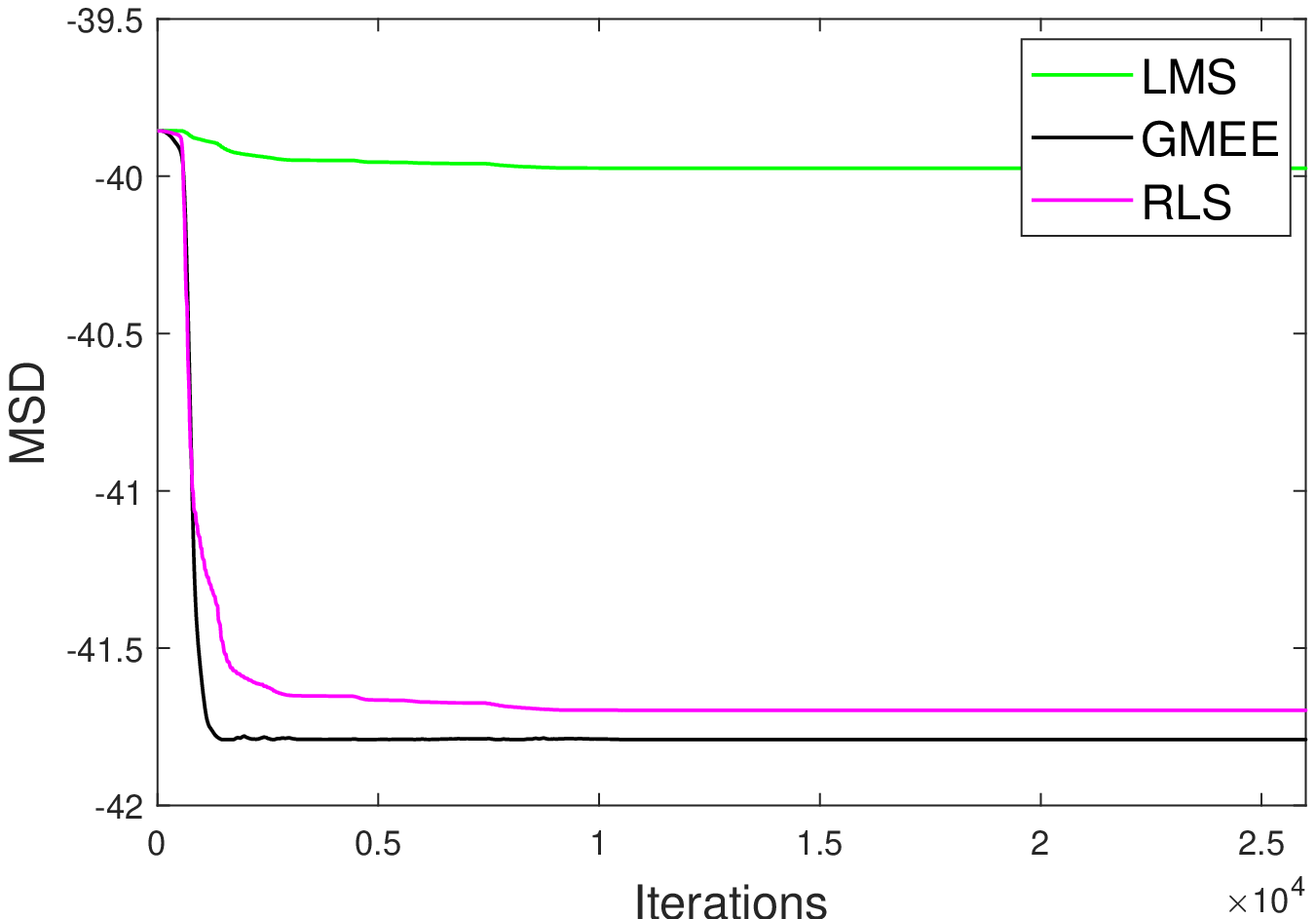}}
\caption{MSDs (dB) of the different algorithms in AEC experiment.} \label{MSDsounds}
\end{figure}

\begin{figure*}[htbp] 
\centering  
\subfigure[The Desired signal.]{
\includegraphics[width=0.47\textwidth]{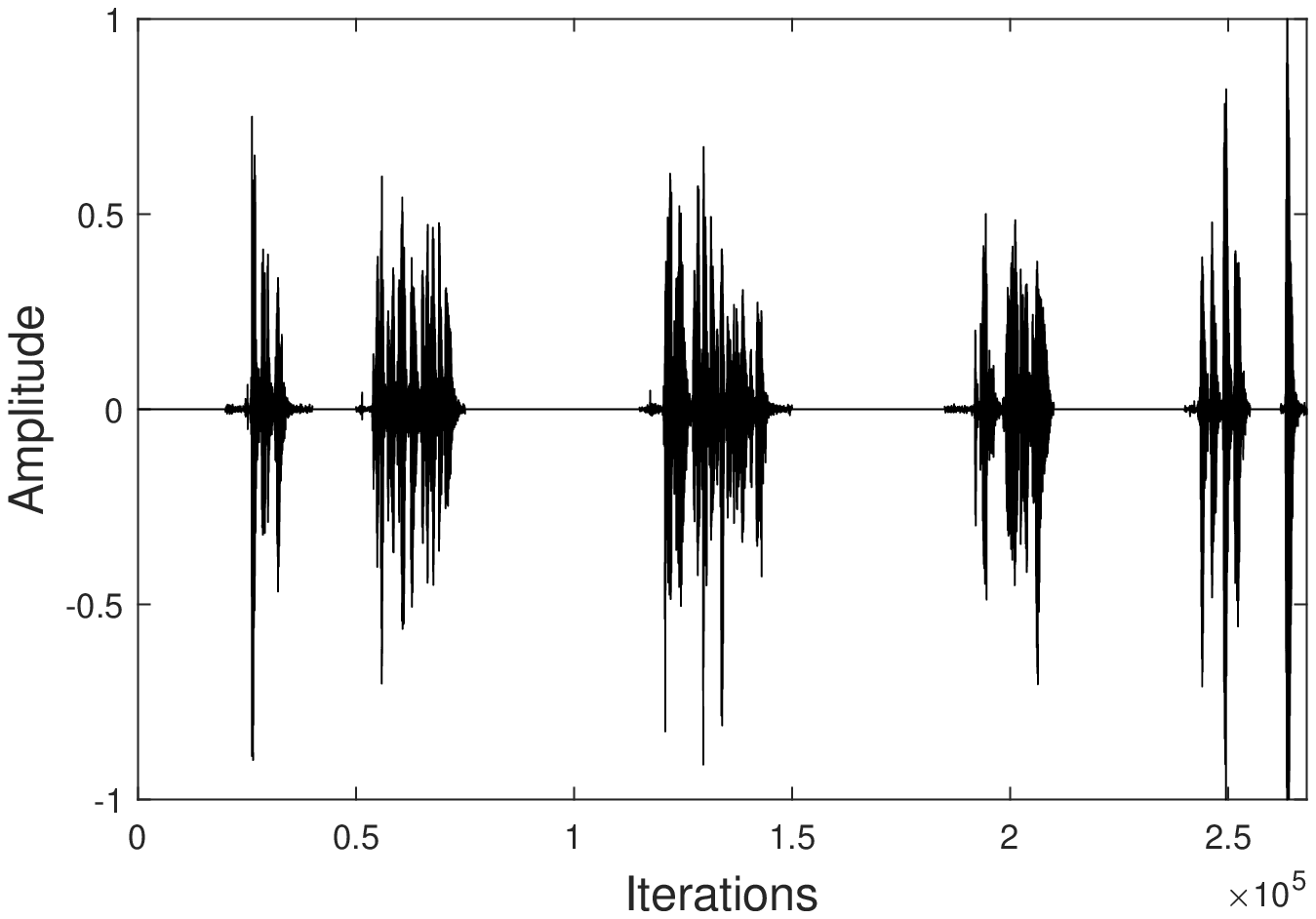}
}
\quad
\subfigure[The echo signal.]{
\includegraphics[width=0.47\textwidth]{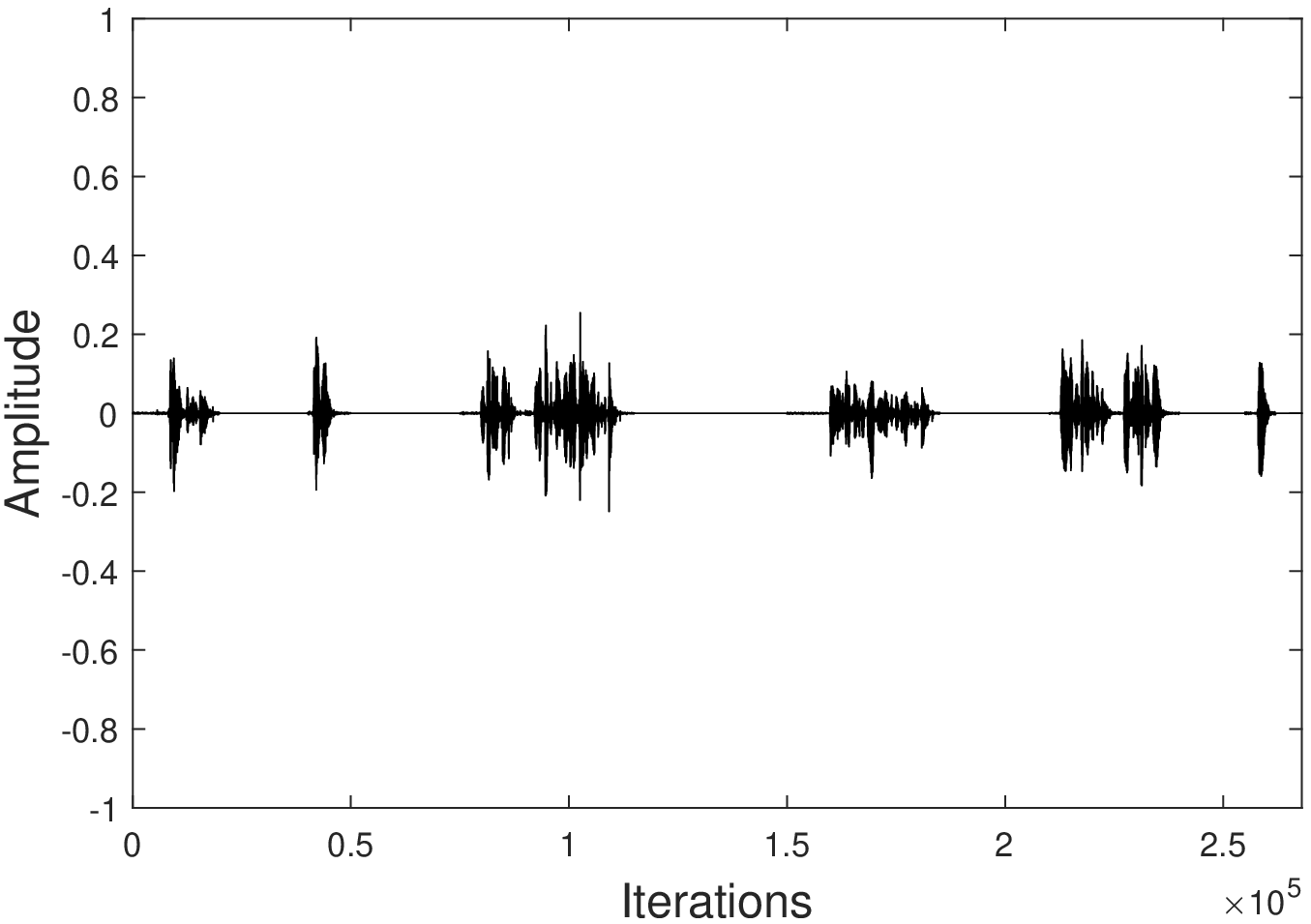}
}
\quad
\subfigure[The microphone signal.]{
\includegraphics[width=0.47\textwidth]{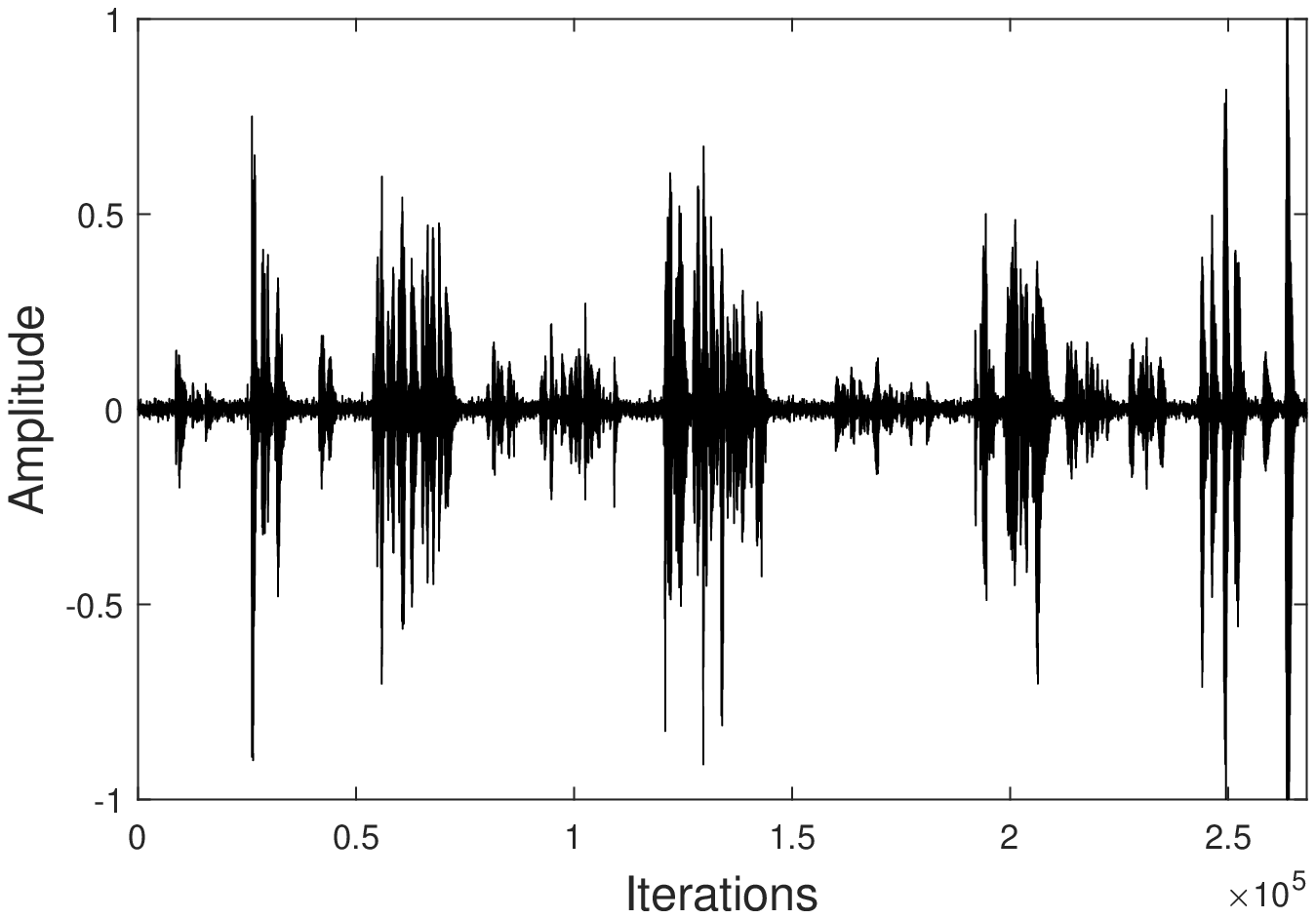}
}
\quad
\subfigure[The signal is processed by the GMEE.]{
\includegraphics[width=0.47\textwidth]{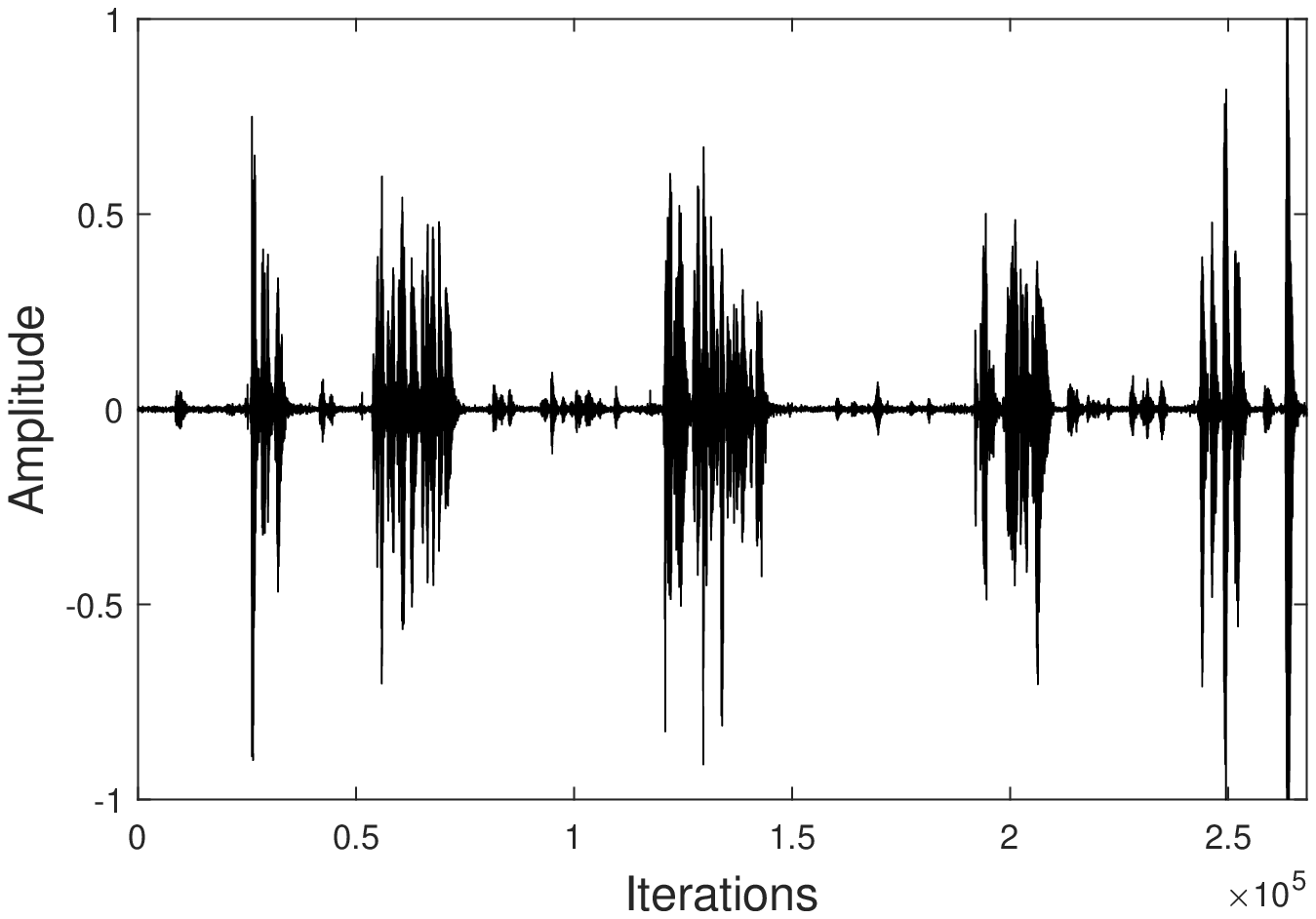}
}
\caption{Sound signal waveforms}\label{sound_waveforms}
\end{figure*}

\begin{figure*}
\centerline{\includegraphics[width=\textwidth]{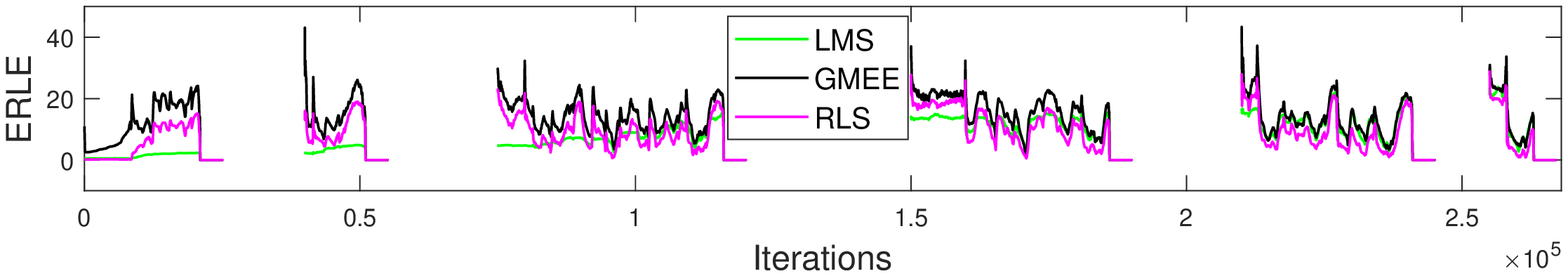}}
\caption{The ERLS of different algorithms.} \label{ERLE}
\end{figure*}

The MSD of the GMEE, LMS, and RLS algorithms are exhibited in fig. \ref{MSDsounds} under this AEC experiment. It can be inferred from fig. \ref{MSDsounds} that the stable MSDs of these algorithms obey rule: ${{\text{GMEE}} < {\text{RLS}} < {\text{LMS}}}$, and the convergence speeds of RLS and GMEE are almost the same. Fig. \ref{sound_waveforms} displays some sound signal waveforms including desired signal, echo signal, microphone signal, and processed signal from GMEE algorithm. It is clearly evident that our proposed algorithm can effectively eliminate echo. In addition, the ERLE of LMS, GMEE, and RLS algorithms are obtained through 1000 Monte-Carlo experiments in fig. \ref{ERLE}. It is obvious that the GMEE algorithm outperform LMS and RLS in respect to ERLE.

\section{Conclusion} \label{Conclusion}
The last few years, error entropy is a similarity measurement method proposed, and the MEE criterion has also been successfully applied to many practical fields. In the available literature, the default kernel function of error entropy is Gaussian kernel function, which is not necessarily the best option. In this paper, the GGD function is utilized as kernel function of error entropy, one can further obtain generalized error entropy and GMEE optimization criterion. In addition, we also propose a novel adaptive filtering called GMEE algorithm using GMEE criterion. Further, the stability, steady-state performance, and computation complexity of the GMEE algorithm are analyzed. Some simulation results indicate that the GMEE adaptive filter algorithm outperforms some existing adaptive filtering algorithms in Gaussian, sub-Gaussian, and super-Gaussian noises respectively. This experiment of applying GMEE to AEC further demonstrates the practicality of the proposed algorithm.

The GMEE optimization criterion with its superior learning performance has been successfully applied to the LMS algorithm. On top of this, there are still some issues that need to be studied. As we know, the RLS algorithm outperforms LMS algorithm, it is natural to develop a new RLS algorithm based on GMEE to improve the performance of the RLS algorithm. This point is the focus of our research in the future.

\section{Acknowledgements} \label{Acknowledgements}
This study was founded by the National Natural Science Foundation of China under Grant 61371182 and 51975107.

\ifCLASSOPTIONcaptionsoff
  \newpage
\fi


\bibliographystyle{IEEEtran}

\begin{IEEEbiography}{Michael Shell}
Biography text here.
\end{IEEEbiography}

\begin{IEEEbiographynophoto}{John Doe}
Biography text here.
\end{IEEEbiographynophoto}

\begin{IEEEbiographynophoto}{Jane Doe}
Biography text here.
\end{IEEEbiographynophoto}

\end{document}